\documentclass[a4paper,fleqn]{cas-dc}

\usepackage{graphicx}
\usepackage{lipsum}
\usepackage[numbers]{natbib}
\usepackage{float}
\usepackage{ulem}
\usepackage{mathtools}
\usepackage{nccmath}
\usepackage{xcolor}
\usepackage{cuted}
\usepackage{amsmath}
\usepackage{multicol}
\usepackage{xcolor}
\usepackage[linesnumbered,ruled,vlined]{algorithm2e}

\SetCommentSty{mycommfont}

\SetKwInput{KwInput}{Input}        
\SetKwInput{KwOutput}{Output}       

\def\tsc#1{\csdef{#1}{\textsc{\lowercase{#1}}\xspace}}
\tsc{WGM}
\tsc{QE}
\tsc{EP}
\tsc{PMS}
\tsc{BEC}
\tsc{DE}

\begin{document}

\let\WriteBookmarks\relax
\def\floatpagepagefraction{1}
\def\textpagefraction{.001}
\shorttitle{RLP Routing}
\shortauthors{Qamar Usman et~al.}

\title [mode = title]{A Reliable Link-adaptive Position-based Routing Protocol for Flying Ad hoc Network}
\author[1]{Qamar Usman}
\ead{engrqamarusman@gmail.com}

\credit{Conceptualization of this study, Methodology, Software}

\author[1]{Omer Chughtai}[orcid=0000-0001-7511-2910]
\cormark[1]
\ead{omer.chughtai@cuiwah.edu.pk}

\author[1]{Nadia Nawaz}
\ead{drnadia@ciitwah.edu.pk}

\author
[1]
{Zeeshan Kaleem}
\ead{zeeshankaleem@gmail.com}

\credit{Data curation, Writing - Original draft preparation}

\address[1]{Department of Electrical and Computer Engineering COMSATS University Islamabad, Wah Campus, Wah Cantonment, Pakistan}
\address[2]{Department of Production Engineering, University of Bremen, Germany}
\address[3]{Institute of Research and Development, Duy Tan University, Da Nang 550000, Vietnam}

\author
[2]
{Kishwer Abdul Khaliq}
\ead{kai@biba.uni-bremen.de}

\author
[3]
{Long D. Nguyen}
\ead{nguyendinhlong1@duytan.edu.vn}


\begin{abstract}
Flying ad hoc network (FANET) provides portable and flexible communication for many applications and possesses several unique design challenges; a key one is the successful delivery of messages to the destination, reliably. For reliable communication, routing plays an important role, which establishes a path between source and destination on the basis of certain criteria. Conventional routing protocols of FANET generally use a minimum hop count criterion to find the best route between source and destination, which results in lower latency with the consideration that there is single source/destination network environment. However, in a network with multiple sources, the minimum hop count routing criterion along with the 1-Hop HELLO messages broadcasted by each node in the network may deteriorate the network performance in terms of high End-to-End (ETE) delay and decrease in the lifetime of the network. This research work proposes a Reliable link-adaptive position-based routing protocol (RLPR) for FANET. It uses relative speed, signal strength, and energy of the nodes along with the geographic distance towards the destination using a forwarding angle. This angle is used to determine the forwarding zone that decreases the undesirable control messages in the network in order to discover the route. RLPR enhances the network performance by selecting those relay nodes which are in the forwarding zone and whose geographic movement is towards the destination. Additionally, RLPR selects the next hop with better energy level and uses signal strength and relative speed of the nodes to achieve high connectivity-level. Based on the performance evaluation performed in the Network simulator (ns-2.35), it has been analysed that RLPR outperforms the Robust and reliable predictive based routing (RARP) and Ad hoc on-demand distance vector (AODV) protocols in different scenarios. The results show that RLPR achieves a 33$\%$ reduction in control messages overhead as compared to RARP and 45$\%$ reduction as compared to AODV. Additionally, RLPR shows a 55$\%$ improvement in the lifetime of the network as compared to RARP and 65$\%$ as compared to AODV. Moreover, the search success rate in RLPR is 16$\%$ better as compared to RARP and 28$\%$ as compared to AODV.

\end{abstract}


\maketitle

\section{Introduction}

Due to the vast technological advancements in mobile computing and in wireless communication, the exchange of information among various devices through wireless means is becoming more popular. Wireless media usually provides low operating- and installation- costs as compared to wired media. Information exchange through wireless medium is generally carried out either through Infrastructure-based or through an infrastructure-less network. The infrastructure-less network is generally referred as an ad hoc network. 
Each and every device in an infrastructure-based wireless network is directly connected to a central entity; and if a device wants to communicate with another device in the network, it needs to communicate through a central node, that is generally known as an access point. A cellular network is one of the common examples of Infrastructure-based network. In such types of network, all mobile devices are connected with the corresponding base station as single-hop fashion. 
One of the major requirements in such types of networks is that each device or node must be under the coverage area of a central entity or base station. However, if a node moves outside the coverage area of a base station then the node might not able to communicate, unless or until there is some graceful handover of connection between the central entities. In an infrastructure-based wireless network, the entire network depends upon a central entity and if a central entity goes down then the entire network collapse.

\begin{figure*}[pos=b]
	\centering
	\includegraphics[scale=0.9]{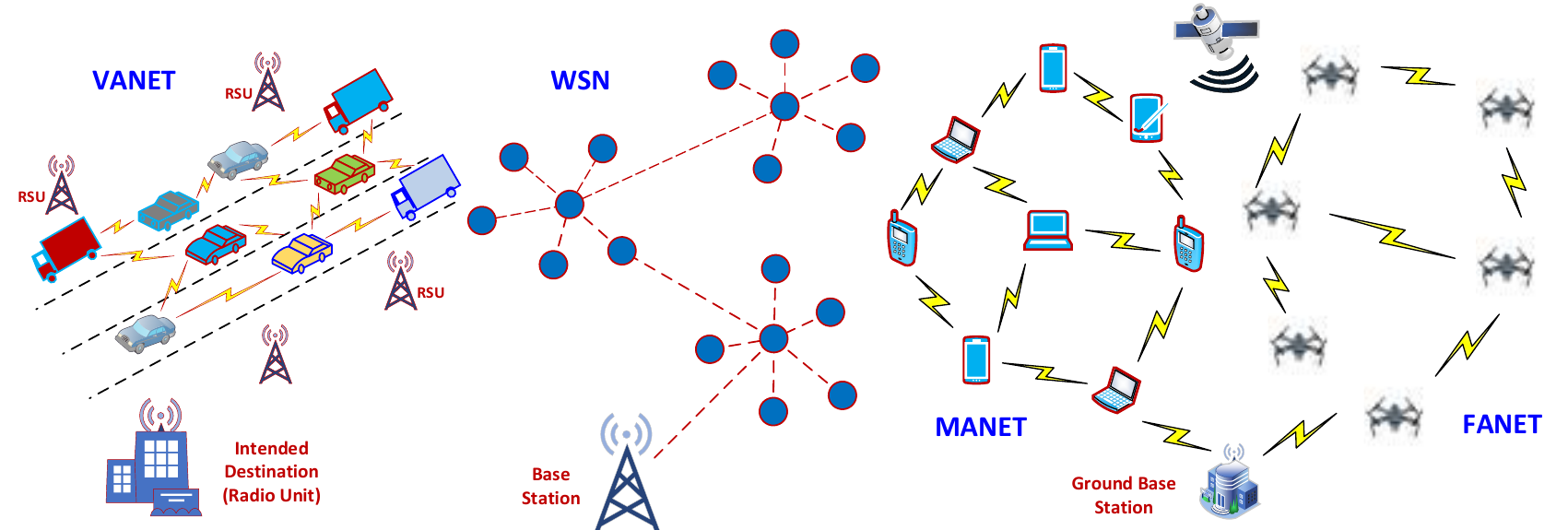}
	\caption{Types of wireless ad hoc networks (infrastructure-less networks)}
	\label{allnetworks}
\end{figure*}

In an ad hoc network with decentralized nature, there is no central entity or base station; therefore, the devices may communicate with each other directly without the intervention of a base station. Here, each node acts as a transceiver or relay with the capabilities of transmitting, receiving, or forwarding the data to its next hop. The devices in an ad hoc network may communicate with each other either in single-hop or multi-hop manner. If a node desires to communicate with a destination and the destination node is not in its vicinity then the intermediate nodes act as relay nodes between source and destination, such type of communication is known as multi-hop communication. Ad hoc network does not depend on pre-established infrastructure and can be deployed where installation of the infrastructure-based network is difficult. Several types of ad hoc network are presented in literature such as Mobile ad hoc network (MANET), Vehicular ad hoc network (VANET), Wireless sensor network (WSN), Wireless mesh network (WMN), Wireless body area network (WBAN), Flying ad hoc network, etc. These networks are classified by their applications and some of the networks are shown in Figure \ref{allnetworks}. FANET can be observed as a special form of ad hoc network. Some of the basic differences of the aforementioned networks are discussed in \cite{bekmezci2013flying} and are tabulated in Table~\ref{tab1}.

\begin{table}[width=.9\linewidth,cols=4,pos=h]
	\caption{Comparison of MANET VANET and FANET.}\label{tab1}
	\begin{tabular*}{\tblwidth}{@{} LLLL@{} }
		\toprule
		\multicolumn{1}{c}{\begin{tabular}[c]{@{}c@{}} Network\\ Characteristics\end{tabular}}~ & \textbf{MANET}~~~ 	& \textbf{VANET}~ 	& \textbf{FANET}~\\
		\midrule
		Nodes Movement 		& Low 		& High 		& Very High 	\\
		Mobility Model 		& Random 	& Regular    & Random    \\
		Nodes Density		& Small		& Large     & Very Small	\\
		Topology Variation   & Low      & High     & Very High   \\
		Energy Consumption~	& Needed    & Not Necessary~ & Required\\
		\bottomrule
	\end{tabular*}
\end{table}

Due to the vast technological advancements in a data communication system, UAV provides new methods of information sharing in military- and civilian-based communication systems~\cite{shakoor2019role}. It provides a flexible way of communication from the last decade with low operating and installation costs. One of the methods to use UAVs for communication is a single UAV-based communication system in which UAV is directly connected with the ground base station. 
In a single UAV-based communication system, if UAV goes down, then apparently the mission cannot be fulfilled. Therefore, instead of using single-UAV-based communication, a group of UAVs is a better alternative, as it may provide enormous benefits compared to a single UAV-based system. One of the advantages of using a multi-UAV-based system is to have a larger monitoring area. 
Moreover, in one of the multi-UAV-based communication systems, where all UAVs are directly linked with ground base-station in a point-to-point fashion; if one of the UAVs among several UAVs goes down, the overall communication can still be preceded; however, if central node expires, the whole communication disturbs.
If any UAV that wants to participate in the communication and is not in the vicinity of the ground base-station, then the communication might not be possible, in that case the intermediate UAVs are required to cooperate and coordinate to forward the traffic to the ground base station.

Several articles have been proposed for UAV-enabled communication using the TCP/IP layering architecture with multi-purpose objectives and resourceful applications. as discussed in the following text. Like for the physical layer, a multi-user communication system has been developed for UAVs to exchange the message through secure means, by using Han-kobayashi signaling (HKS) \cite{1stPaper}. Similarly, for disaster-relief, a mobile crowd sensing based stochastic system has been proposed \cite{2ndPaper} for dynamic environments using social-aware UAVs to show the average payoff of the Mobile crowd sensing (MCS) and UAV with respect to the number of subregions. It showed that payoff increases with the increase in the number of subregions becuase of increasing the utilization rate of the UAV. In analogous to this, software defined network, cloudlet, and the radio access network layers based public disaster-resilient edge architecture for delay-sensitive communication has been proposed in \cite{3rdPaper}. It has been compared with the conventional computing that uses centralized approach, and the results showed better performance in terms of delay and energy consumption.

\begin{figure*}[pos=!t]
	\centering
	\includegraphics[scale=0.80]{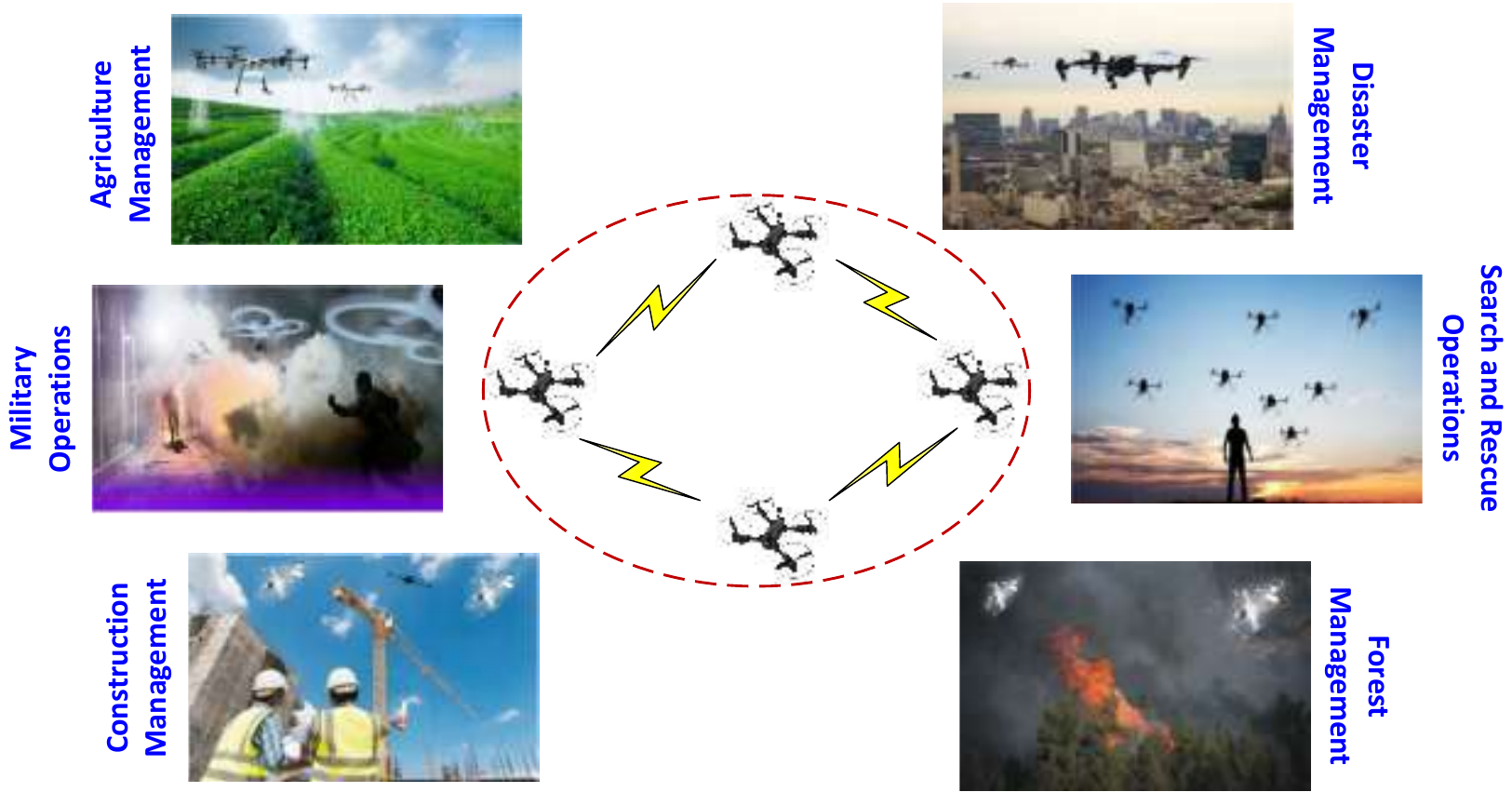}
	\caption{The practical application scenarios of flying ad hoc network}
	\label{fapp}
\end{figure*}

Numerous other applications have been developed that show the usability of FANET. Some of the applications are shown in Figure~\ref{fapp}. UAVs are usually placed in the air with the ability to communicate with each other so that they can work together to perform a specific or assigned task. As discussed earlier that FANET provides new methods of information sharing for the military- and civilian-based communication systems. In a military operation, FANET can be used for border surveillance and monitoring. Additionally, drones (UAVs) are usually equipped with cameras and they may continuously measure the activity of the enemy on the border and send updated reports to the base station.
In search-and-rescue applications, UAVs may be used to access the inaccessible areas. The network of UAVs can be used by mobile phones to find probable survivors or it can be used in disaster situations like an earthquake. FANET can provide mobile services to the users, if existing cellular system goes down. Traffic monitoring is another application of FANET, in which UAVs can notice and report traffic crashes, easily. FANET also plays a vital role in the agriculture field, it can be used for spraying pesticides in crops. FANET can also be deployed in forest to monitor any disaster situation.
Despite numerous advantages, FANET encounters several unique challenges. One of the Unique challenge of FANET is successful reliable delivery of data to destination with minimum End-to-End delay.

The problem of unreliable data delivery and high ETE delay in FANET is due to the inappropriate selection of path or next-hop node in the network. Because of this, the packet-loss rate would increase, thereby, leading to an increase in the control messages overhead, which would degrade the overall network performance. Furthermore, in majority of the techniques, ETE delay is reduced by selecting a route with a minimum number of hops. Such techniques reduce average throughput and increase energy consumption in the network.
Because of a wide variety of applications, high energy efficiency and low installation cost; FANET became a major research motivation for governments, defense, industries, and academic institutions. In~\cite{bekmezci2015flying}, a Testbed has been designed and developed for FANET with low-cost and with easily available hardware; therefore, the desired objectives based on the constraints can be implemented for FANET within a limited budget. The cooperation among UAVs in multi-hop communication can increase the performance of the entire network. Swarms of UAVs increase the network scalability, as higher numbers of UAVs complete a task more efficiently. Additionally, the mission can be completed faster and precisely in a timely manner with the consideration of the environment and topological structure. In most of the applications, performance is directly associated with the number of UAVs, like in search-and-rescue operations~\cite{waharte2010supporting}. However, the successful deployment of FANET consists of a number of important components, a key one of which is the successful delivery of a message to a destination. Data communication between nodes is one of the crucial challenges in FANET. As described earlier, the communication between source and destination is usually carried out using single-hop or multi-hop fashion. Information sharing with the UAV, which is within the vicinity of a source UAV is usually done through a single-hop communication. In contrary to this, the information sharing with the UAV which is not within the vicinity of a particular source UAV is usually done through multi-hop communication. Intermediate nodes retransmit the message until the message reaches a destination. Furthermore, data communication in multi-hop communication is usually done through flooding or routing. 

In flooding each node in the network rebroadcasts the received information once; until the it reaches to destination. Considering the basic phenomenon of flooding, one of the disadvantages is the redundant packets that are received by the node, in accordance with node density. Additionally, flooding creates high traffic in the network, which increases the energy consumption that leads to a decrease in network lifetime. Conversely, routing procedure establishes a route between source and destination and only those nodes are allowed to forward a message, which knows the route to a destination. All the other nodes discard the message. This mechanism reduces duplicate message arrival at a destination. Moreover, the routing path is created on the basis of certain criteria, such as received signal strength, power consumption, buffer utilization, remaining energy, etc.

Traditionally routing procedures use a minimum hop-count criterion to discover a route between source and destination. However, it is quite possible that in a single source and destination network environment, the minimum hop-count criterion may result to have a better performance in terms of ETE delay. However, in a network with multiple sources, this criterion may deteriorate the network performance in terms of high ETE delay, unreliability, and may increase the control messages overhead along with the energy consumption. So FANET requires a routing protocol that rapidly accepts topology variations in the network and may able to forward the traffic reliably with reduced ETE delay and with a smaller number of control messages in the network. By considering the above-stated problems, the energy-efficient route selection mechanism is required that can enhance the network lifetime and also reduce the control messages overhead. Therefore, the aim of this research work is to design and develop a lightweight, reliable, and energy-efficient routing protocol for FANET. In this context, the specific research objective is to design and develop a routing protocol that allows the nodes along the ETE route to select an appropriate next-hop node based on the forwarding zone towards the destination in order,
		\begin{enumerate}[1]
			\item To reduce the control signaling overhead.
			\item To increase the network lifetime.
		\item 	To reduce the number of messages required to discover the destination per unit time that is the search success rate.
		\end{enumerate}

Moreover, the goal of this research work is to design and develop a reliable link adaptive position-based routing protocol for FANET. Evaluation of the proposed routing protocol is carried out through simulation and is compared with the state-of-the-art routing protocols of FANET. 
	The algorithm of the proposed work is developed and simulated using a network simulator (ns-2.35). The following list shows the major contribution of this research work:

\begin{itemize}
	\item A reliable link-adaptive position-based routing protocol is designed for FANET that can work in a highly dynamic environment and enhances the network performance in terms of high connectivity-level with a minimum control overhead.
	\item An on-demand strategy of reactive routing protocols is used with the assumption that each node or UAV in the network knows about its position and the position of the destination and then the path between source and destination is created on the basis of composite metrics using geographic distance and relative speed.
	\item Forwarding zone of each node is determined to identify the front relative nodes and then only the selected front relative are allowed to rebroadcast the received packet. This reduces the number of control messages in the network. Here, only those UAVs are selected as a front relative nodes that have better energy level.
	\item Each UAV selects the next suitable UAV with a better connectivity-level with respect to the relative speed and signal strength. Additionally, for reliable communication, next hop is selected on the basis of link-characteristic metric by using the geographical position of the node that is closest to destination.
	\item An interval (zoom out) is initialized that is used to trigger the node to discover the next hop node based on the link-layer dis-connectivity, during the course of data communication.
\end{itemize}

The rest of the paper is organized in such a way that, Section \ref{literature} discusses the literature review. Section \ref{proposed} thoroughly elaborates on the proposed routing mechanism along with the system model. Next-to-Follow is Section \ref{results}, that illustrates the simulation and results. Finally, Section \ref{conclusion} discusses the conclusion and future work.

\section{
	Literature review}
In this section existing routing protocols of FANET and their objectives are discussed, and in the end, this section gives an overview for the design challenges of routing protocols in FANET.
\label{literature}

\subsection{\textbf{FANET routing protocols}}
\begin{figure*}[pos=t]
	\centering
	\includegraphics[scale=0.69]{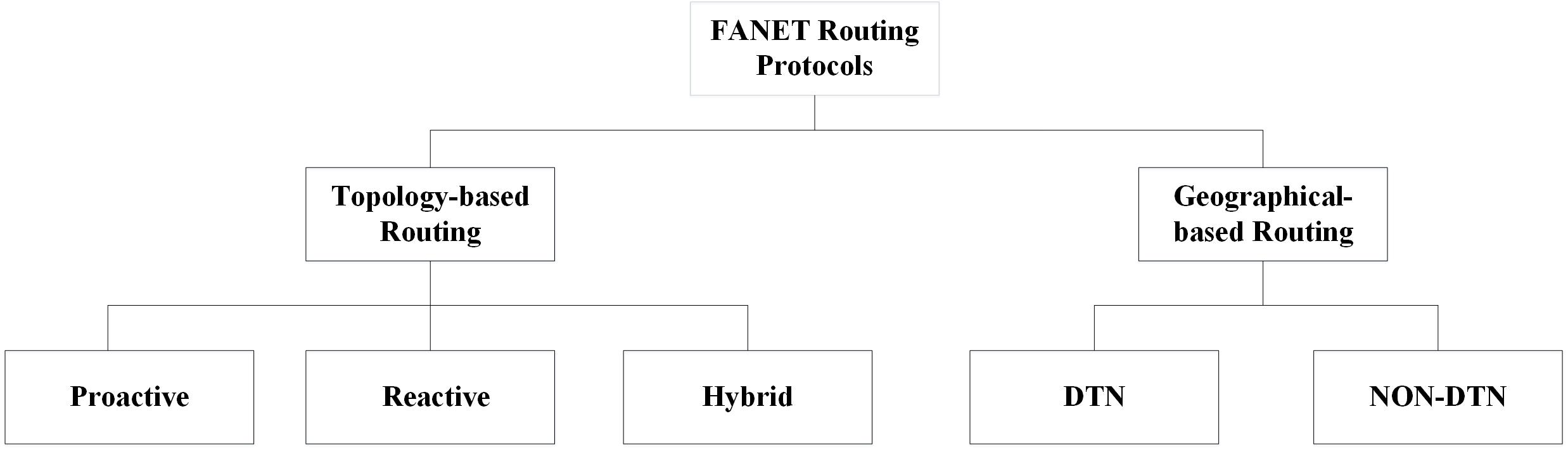}
	\caption{Classification of Flying ad hoc Network Routing Protocols}
	\label{rp}
\end{figure*}

Routing protocols of FANET are generally categorized into two main groups~\cite{arafat2019routing}. The first one is the Topology-based routing and the second is the Geographical-based routing, as presented in Figure~\ref{rp}. The illustration of these categories with reference to their state of the art routing protocols is discussed in the following text.

\subsubsection{\textbf{Topology-based routing protocols}}
Topology-based routing protocols need topological information of nodes and the associated links along with the orientation to create and maintain the routing path. Nodes generally interchange the topology information with other nodes and on the basis of topology information, the routing table is constructed to record the information about next hop. Topology-based routing is further categorized into three categories such as; proactive, reactive, and hybrid. 

Proactive routing protocols are also called table-driven protocols. In these types of routing protocols, nodes periodically exchange HELLO messages or 1-hop messages based on the vicinity in the network with the neighboring nodes. With the shared information, the routing table is created and maintained. Furthermore, the routing table is updated periodically and each node stores the received information and compares the entries with the already stored one. Through this process, the complete topology information of the network is available on each node. If there is any change in the network, then the update is gain shared with all the nodes. Due to the high mobility in FANET, periodic updates for the routing table are not optimal. One the other hand, one of the main benefits of proactive routing protocols is that the nodes store recent information about the routes. But due to the maintenance of the routing information at each node, overhead as signaling, increases in FANET because of the high mobility and therefore, proactive routing is not suitable for such networks.

However different proactive routing protocols were proposed for FANET by different authors. In~\cite{alshabtat2010low} authors proposed Directional optimized link state routing (DOLSR) over a well-known proactive routing protocol that is the Optimized link state routing (OLSR). 
In DOLSR, UAVs are equipped with a directional antenna. If a node has a data packet for transmission, it first calculates distance towards the destination node. In case of greater distance than the ${\raise0.7ex\hbox{${\mathop D\nolimits_{\max } }$} \!\mathord{\left/
		{\vphantom {{\mathop D\nolimits_{\max } } 2}}\right.\kern-\nulldelimiterspace}
	\!\lower0.7ex\hbox{$2$}}$ (where $\mathop D\nolimits_{\max }$ is extreme range of directional antenna), then DOLSR uses directional antenna for data transmission. If distance is less than ${\raise0.7ex\hbox{${\mathop D\nolimits_{\max } }$} \!\mathord{\left/
			{\vphantom {{\mathop D\nolimits_{\max } } 2}}\right.\kern-\nulldelimiterspace}
		\!\lower0.7ex\hbox{$2$}}$ then a node applies the conventional OLSR protocol settings and the omni-directional antenna is used. Using DOLSR routing protocol, fewer multipoint relays are selected as compared to OLSR and that would attain performance improvement in terms of lower communication overhead as well as minimum ETE delay.

In~\cite{rosati2013speed} predictive-OLSR (POLSR) protocol is proposed by authors. In this protocol geographical positions of the nodes are exchanged through the HELLO packets. On the basis of geographical positions, nodes know the location of its neighbors and the Expected transmission count (ETX) metric~\cite{ni2008performance} is calculated through which the relative speed between two nodes is obtained. Through such information, the quality of the communication link is evaluated and nodes with higher link-quality are selected as the next forwarding node in the network. Another proactive routing protocol that is the Link-quality traffic load-aware optimized link state routing (LTA-OLSR) is proposed in~\cite{pu2018link}. The objective of LTA-OLSR is to deliver reliable data transmission for FANET. In this protocol, link awareness was considered to distinguish link qualities among neighbor nodes by taking information of Received signal strength indicator (RSSI).

In~\cite{xie2018enhanced}, pengcheng xie extends the functioning of OLSR routing protocol and proposed OLSR-ETX routing protocol for Ocean FANETs, that is built on nodes' link termination time and remaining energy. For the selection of Multi-point relay (MPR) nodes, a new metric using ETX and residual energy is used. Here, location information is used to estimate the ETX. Here, that node is selected as MPR whose ETX is smaller than one and whose energy is greater than the threshold energy. Mobility and load aware OLSR (ML-OLSR) protocol is suggested by~\cite{6913628}. ML-OLSR avoids choosing a high-speed node as MPR because of the usage of relative speed and the location of neighboring nodes. To reduce packets' interference among neighboring nodes, the packet load of all UAVs is also taken into account to find steady routes with less congestion. In a proactive routing protocol, the routing table must be maintained each time, whenever there is any change occur in the network. So huge Bandwidth is required to maintain routing information. In contrary to this; to minimize the usage of bandwidth consumption in the network, reactive routing can be utilized. 

In reactive routing, routes between the nodes are determined only when the host wanted to forward packets. Such routing strategies are suitable solutions for highly dynamic FANET. However, such strategies may suffer from high latency, due to the time it takes to discover the route. Numerous reactive routing protocols are proposed for FANET. 
In~\cite{hong2019tarcs}, a Topology-aware routing protocol with choosing scheme (TARCS) is proposed that builds the effects of periodic topology change awareness (PTCA) and Adaptive route choosing scheme (ARCS). PTCA precisely measures topology change and decides the movement. ARCS selects an appropriate routing for current motion and to certify that network performance is affected as little as possible. In~\cite{choi2018geolocation}, authors suggested a Geolocation-based multi-hop routing protocol (GLMHRP). The purpose of this protocol is to conserve robust connectivity between ground stations and UAVs. Each node shares its position, speed, and direction to its neighbors, continuously. From this information, every node selects its forwarder node. The metrics of selecting the next nodes are based on the greedy forwarding technique. Link stability estimation based preemptive routing (LEPR) is presented in~\cite{li2017lepr}. In this protocol, the geographic position of the node is used to predict node's mobility and link-quality between the nodes. For route discovery, all disjoint paths are rejected and only that path is selected which has fully connected nodes. 

Hybrid routing protocols are a combination of reactive and proactive routing. In~\cite{liu2008clustering}, a Hybrid routing protocol based on clusters (HRC) is presented. In this protocol, proactive routing is used for intra-cluster and reactive routing is used for inter-cluster data transmission. Because of the high mobility in FANET, maintaining and storing the routing tables using a proactive routing procedure is not feasible. In contrary to this, finding a routing path every time in reactive routing is a costly procedure. A Mobility prediction clustering algorithm (MPCA) was proposed in~\cite{zang2011mobility}. Where a GPS-based data is used to compute the Link expiration time (LET) between two UAVs. The UAV with the largest permanent neighbor has the highest linking degree to be represented as a cluster head. Furthermore, MPCA increases network scalability. The main problem of MPCA is that it is not energy-efficient in a highly dynamic network environment. In~\cite{aadil2018energy}, an energy-aware and link-based routing is discussed. In this protocol, authors have applied optimization to determine the routing path with the goal to save the energy of the UAVs by means of controlling their transmission range to efficiently clustering the whole network.

\subsubsection{\textbf{Geographical-based routing protocols}}

To solve various problems of topology-based routing protocol, geographical-based routing is proposed, in which the routing path is established by predicting the positions of the nodes in the network. Each node defines its position using Global positioning system (GPS) and the nodes using geographic routing, deliver a message over multiple hops by knowing nodes' position. Decisions of the routes are neither built on network addresses nor through maintaining routing tables. Therefore, routing decision in geographical routing generally does not take place using routing tables. Each node selects a next-hop node from its neighboring nodes whose position is closest to the destination. The position of the neighbors is obtained by periodically exchanging beacon packets with the neighboring nodes. Geographical routing protocols capable to work in highly dynamic situations like used in FANET. The geographical-based routing is classified into two sub categories; Delay tolerant networks protocols (DTN) and Non-delay tolerant networks (non-DTN).

\noindent \textbf{1. Delay tolerant networks protocols (DTN)}

In Delay tolerant network (DTN) techniques, nodes store data for some time until another node meets, by moving into the vicinity. These methods are designed to resolve technical problems of networks that are usually suffered from regular disconnections, because of high mobility. Additionally, nodes are required to save data until they meet another forwarding node (s). Although, such protocols decrease the packet loss, at the same time these protocols increase the delay in the network. A geographical and delay-tolerant Location-aware routing for opportunistic delay tolerant (LAROD) network is presented in~\cite{kuiper2010geographical}. LAROD is built on store-carry-and-forward and with a greedy forwarding mechanism to forward data packets to the destination. Each node is required to broadcast a data packet to its neighboring node. Neighboring nodes after receiving the data packet, hold it for some time and start a timer, accordingly. Any intermediate that has the least expiry time, rebroadcast the data packet. Other nodes will just overhear the rebroadcast transmission. This prevents the network to forward duplicate packet transmission. All the nodes in the network use the same procedure until the data packet reached to the destination.

In~\cite{hyeon2010new}, a Geographic routing protocol for aircraft ad hoc networks (GRAA) is proposed. The decision of the path is made locally at each and every intermediate node. To determine the selection of the next hop, every node in the network predicts the location and the speed of the neighbors along with the destination. Firstly, the source UAV computes the predictable location of a destination and afterward with a time period $t$, based on its present location, its speed is calculated. After this, it computes the predictable location of all its neighbors. Node that has the closest location to the destination is selected as next-hop among multiple nodes, based on the node density. Authors proposed Robust and reliable predictive based routing (RARP) in~\cite{gankhuyag2017robust}. The aim of this protocol is to achieve a high connectivity level among the nodes in a highly dynamic network. Each next relay node is selected by using the optimum value of the determined composite metric, which is based on expected connection time and the time duration during which node is in the vicinity of the previous node and with the minimum hop count. The risk value of the network is also computed, which is based on the operational requirements and environmental conditions. The destination node receives multiple route request packets and waits for a certain amount of time and then sends a route reply to that node which has a better utility function. RARP uses the omni-directional antenna to exchange the route request packet between nodes to discover a routing path and the data-packets are sent using a directional antenna. During data forwarding, each node also piggybacks its fresh location and speed. Each and every receiving node updates the received information and maintain the forwarding table. From simulation results, it can easily verify that RAPR routing increases route setup success rate by achieving the high connectivity-level between the nodes.

As examined previously, proactive-based routing depends on routing tables that must be updated periodically, even if there is no data packet for transmission. This introduces a large number of control messages which increases the overhead and also consumes a lot of energy. On the other side, reactive routing protocols update the routing tables, on-demand. Comparatively, reactive routing is more appropriate in an extremely dynamic environment, when compared with the proactive routing. But reactive routing protocols have extensive transmission delay. On the other side, hybrid routing protocols give better performance as compared to the proactive and reactive routing protocols. This is because of the usage of the benefits of both techniques. The performance of geographical-based routing strongly depends on the correct prediction of nodes' position. This minimizes the overhead of the network and enhances the scalability of the network. The comparative analysis of FANET routing protocols is shown in Table~\ref{Comparsion}.

\noindent \textbf{2. Non-delay tolerant networks protocols (non-DTN)}

\begin{table*}[pos=hbt]
	\centering
	\caption{Comparative Analysis of FANET Routing Protocols}
	\label{Comparsion}
	\begin{tabular}{|l|l|l|l|l|l|l|l|l|l|l|l|l|l|l|}
		\hline
		\multicolumn{1}{|c|}{\begin{tabular}[c]{@{}c@{}}Routing \\ Protocols\end{tabular}} & \multicolumn{3}{c|}{\begin{tabular}[c]{@{}c@{}}Topology-based\\ Routing Protocols\end{tabular}} & \multicolumn{2}{c|}{\begin{tabular}[c]{@{}c@{}}Geographical-based\\ Routing Protocols\end{tabular}} & \multicolumn{5}{c|}{\begin{tabular}[c]{@{}c@{}}Design Protocol Goals\end{tabular}}                                       & \multicolumn{4}{c|}{\begin{tabular}[c]{@{}c@{}}Routing Metrics\end{tabular}} \\ \hline
		& \rotatebox{90}{    Proactive }                & \rotatebox{90}{Reactive }           & \rotatebox{90}{Hybrid }          & \multicolumn{1}{c|}{\rotatebox{90}{DTN } }               &\rotatebox{90}{NON-DTN }                      & \multicolumn{1}{c|}{\rotatebox{90}{Packet delivery ratio } } & \multicolumn{1}{c|}{\rotatebox{90}{Low Latency }} & \multicolumn{1}{c|}{\rotatebox{90}{BW Efficient} } & \multicolumn{1}{c|}{\rotatebox{90}{Low packet overhead }} & \rotatebox{90}{Scalability} & \rotatebox{90}{ETX}   & \rotatebox{90}{Minimum hop count}    & \rotatebox{90}{Optimized Link}  & \rotatebox{90}{Path life Time}  \\ \hline
		D-OLSR                                       & \multicolumn{1}{c|}{$$\checkmark$$}          &              &              & \multicolumn{1}{c|}{}                & \multicolumn{1}{c|}{}              &                      &           $$\checkmark$$   &                  &                     &       &      &             $$\checkmark$$&           &           \\ \hline
		P-OLSR                                       &                     $$\checkmark$$ &               &              &                           &                         $$\checkmark$$&                     $$\checkmark$$ &             $$\checkmark$$  &                  &                     &       &    $$\checkmark$$  &             &           &           \\ \hline
		M-OLSR                                       &  $$\checkmark$$                   &               &              &                           &                         &                      $$\checkmark$$&              $$\checkmark$$ &                  &                     &       &      &             $$\checkmark$$&           &           \\ \hline
		OLSR-ETX                                      & $$\checkmark$$                     &               &              &                           &                         &   $$\checkmark$$                    &  $$\checkmark$$             &                  &                     &       &$$\checkmark$$      &             &           &           \\ \hline
		TS-AODV                                      &                     & $$\checkmark$$              &              &                           &                         & $$\checkmark$$                      &              &                 $$\checkmark$$ &                     &       &   $$\checkmark$$   &             &           &           \\ \hline
		RTORA                                       &                     & $$\checkmark$$              &              &                           &                         &                      &              &  $$\checkmark$$                &    $$\checkmark$$                 &       &      &            $$\checkmark$$ &           &           \\ \hline
		HRC                                        &                     &               &      $$\checkmark$$        &                           &                         &                      &              &    $$\checkmark$$              &     $$\checkmark$$                 &       &      & $$\checkmark$$            &           &           \\ \hline
		MPCA                                        &                     &               &          $$\checkmark$$    &                           &                         &                      &              &                  &                     &      $$\checkmark$$ &      &             &         $$\checkmark$$  &           \\ \hline
		GPMOR                                       &                     &               &              &                           &    $$\checkmark$$                      &                     $$\checkmark$$  &              &                  &                     & $$\checkmark$$      &      &    $$\checkmark$$          &           &           \\ \hline
		GPSR                                        &                     &               &              &                           &                        $$\checkmark$$  &                      &              &                  & $$\checkmark$$                     & $$\checkmark$$      &      &            $$\checkmark$$ &           &           \\ \hline
		MPGR                                        &                     &               $$\checkmark$$ &              &                           &                         $$\checkmark$$ &     $$\checkmark$$                  & $$\checkmark$$              &                  &  $$\checkmark$$                    &       $$\checkmark$$ &      &             &           &           \\ \hline
		RGR                                        &                     &               &              &                           & $$\checkmark$$                         &                      $$\checkmark$$ &              $$\checkmark$$ &                  &                     $$\checkmark$$ &       &      &  $$\checkmark$$            &           &           \\ \hline
		ARPAM                                       &                     &  $$\checkmark$$              &              &                           & $$\checkmark$$                         &                      &              &  $$\checkmark$$                &                     &       &      &         $$\checkmark$$    &           &           \\ \hline
		LAROD                                       &                     &$$\checkmark$$               &              & $$\checkmark$$                           &                         &                      $$\checkmark$$ &             $$\checkmark$$  & $$\checkmark$$                 &                     &       &      &             &      $$\checkmark$$     &           \\ \hline
		GRAA                                        &                     &               &              &                           $$\checkmark$$ &                         & $$\checkmark$$                      &              &                  &                     &       $$\checkmark$$ &     $$\checkmark$$  &             &           &           \\ \hline
		RARP                                        &                     &   $$\checkmark$$             &              &                           $$\checkmark$$ &                         &                      $$\checkmark$$&              &                  &  $$\checkmark$$                   &       &      &             &           & $$\checkmark$$          \\ \hline
		
	\end{tabular}
\end{table*}

The key objective of non-DTN-based protocols is to transmit the data packet to the destination as fast as possible. Non-DTN-based applications may use a beacon-based or beacon-less strategy. In~\cite{lin2012novel}, the authors proposed Geographic position mobility oriented routing (GPMOR) using a non-DTN geographical routing protocol. In GPMOR, Gauss-Markov mobility model~\cite{broyles2010design} is used to predict position of node. Each and every UAV is required to determine its geographical position through GPS. Additionally, all UAVs in the network periodically broadcast their position and velocity to their neighboring nodes through HELLO beacon packets. If a node has data to send to a destination, it uses the position of a destination and the neighbors, then it calculates the distance between the destination with respect to each neighbor node. The node with the shortest distance from the destination is selected as a next-hop. In~\cite{karp2000gpsr} another geographic routing protocol as Global positioning stateless routing (GPSR) is presented. Here, if a node needs to transmit data, it first predicts the location of the intended destination. Routing decision on each and every node is constructed based on the location of the destination node and then all forwarding nodes are required to select the next hop, which is nearer to the destination to forward the data packet. Mobility prediction based geographic routing (MPGR) is proposed in~\cite{lin2012geographic}. MPGR protocol has the same principle as used in GPSR. Furthermore, MPGR predicts mobility based on the gaussian distribution function. When a UAV has a data packet for transmission, it broadcasts a Neighbor discovery (ND) packet to find the optimal next forwarding UAV on the basis of information, which is included in the reply packet. In~\cite{lin2012geographic}, a Reactive greedy routing (RGR) is proposed. In RGR, the path is established using AODV protocol and the source node starts sending data towards the destination, after path discovery. Due to the high mobility, the path is established between the source and destination, which may be disconnected. In such a situation, RGR switches to a Greedy geographic forwarding mode (GGF) and the nodes forward the data packets to a node which is nearest to the destination.

In~\cite{iordanakis2006ad}, Ad hoc routing protocol for aeronautical MANETs (ARPAM) is proposed. Its working principle is the same as used by AODV. However, geographical positions of the UAVs are used to select the shortest path along with the minimum hop count, through the dissemination of RREQ packets in the network. In ARPAM, the velocity vector and the geographical location of node are used. The intermediate nodes use this information to predict the position of a UAV and select the shortest path between source and destination.

\subsection{\textbf{Design challenges for new reliable routing protocol}}
Considering the network challenges of data transmission in FANET, as discussed in Table \ref{Comparsion}, there is a need to design and develop a routing strategy that can work in a highly dynamic environment as used in FANET. The majority of the routing protocols in the literature, establish a route between source and destination, using a single criterion such as the minimum number of hops. Using the minimum number of hops between source and destination may results in better network performance in terms of reduced ETE delay, but with the assumption of a single source and destination network environment. However, with multiple sources, the minimum number of hops as a routing criterion may increase the ETE delay because of the bottleneck at some intermediate node in the network. This is because, probably the same path might be selected by most of the sources in a multi-source environment, which may deteriorate network performance in terms of high ETE delay, energy consumption, packet loss probability, and reduced reliability.

Moreover, the reactive routing strategy is more suitable in FANET to establish a route between the source and destination. With reference to the literature review discussed in Section \ref{literature}, it is concluded that when a source node wants to establish a routing path, it broadcasts a route request to all its 1-Hop neighbors, like used in conventional AODV protocol. Where all 1-Hop neighbors rebroadcast the route request until a path for the desired destination is found or the actual destination is discovered. This increases the control overhead in finding the path, however, a path may be selected with poor link-quality or those nodes are being the part of the path with remaining energy less than the threshold. So, from all these challenges, the neighbor selection is one of the crucial challenges in FANET. Therefore, the reliable delivery of a message to the destination is dependent upon next-hop selection. The next-hop must be selected on the basis of certain metrics to achieve high reliability in the network. Considering these facts, a reliable link-adaptive and position-based routing protocol is proposed in this work, which enhances the network performance as discussed in detail in Section \ref{proposed}.

\begin{figure*}[pos=!b]
	\centering
	\includegraphics[scale=0.9]{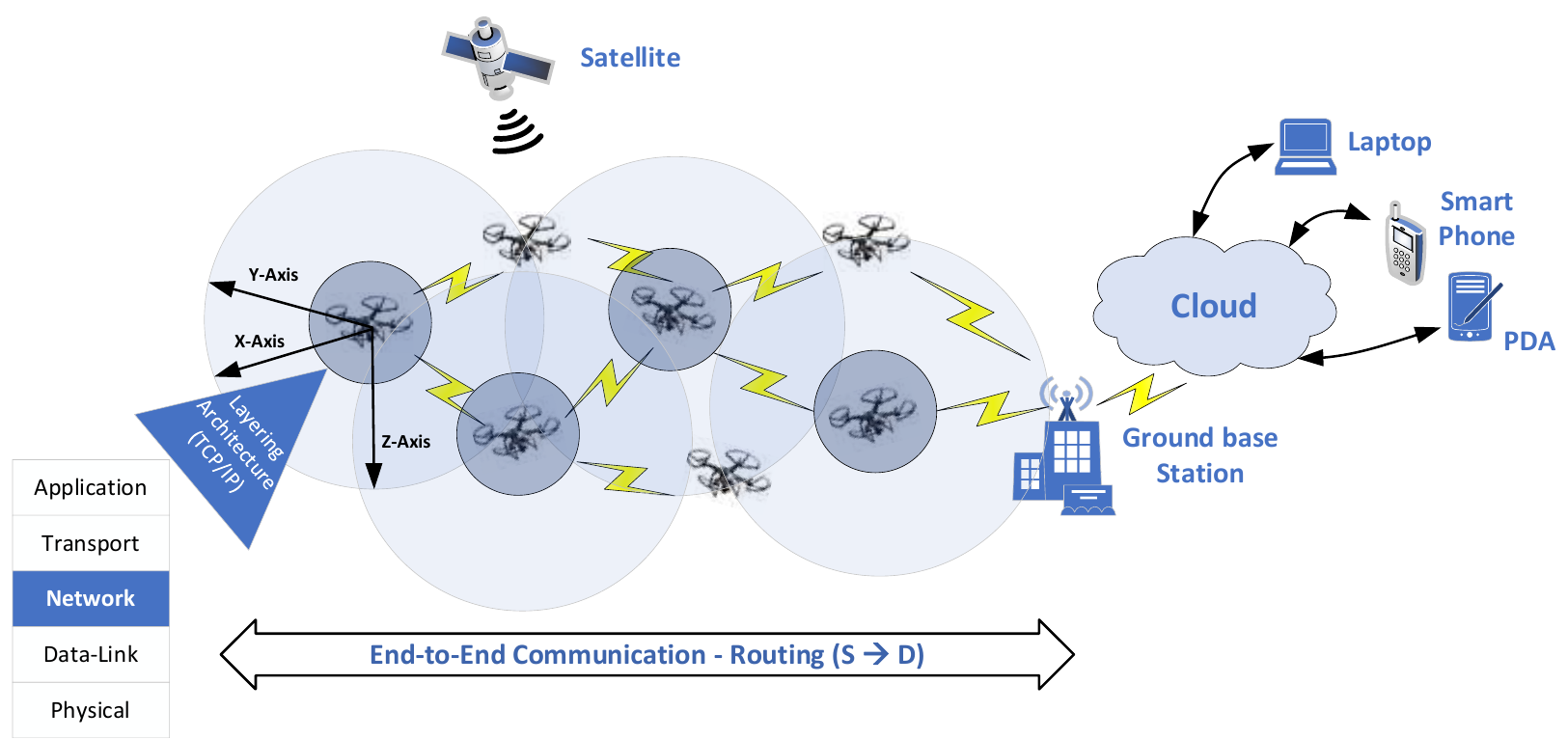}
	\caption{System Model for RLPR routing Protocol}
	\label{sm}
\end{figure*}

\section{\textbf{Proposed reliable link-adaptive position-based routing protocol for FANET}} 
\label{proposed}
This section provides a description of the proposed routing protocol. It includes those routing metrics that directly affect links and nodes deployed in the network. Furthermore, a complete working principle of the proposed routing protocol is presented. 

\subsection{\textbf{System model}}
A system model based on the assumptions used in designing the proposed routing protocols is discussed in the following text. It is assumed that each node or UAV knows the position of the destination in order to estimate the direction towards the destination.  Figure~\ref{sm} represents a system model for the prosed work in FANET, where a number of UAVs creates an ad hoc network with each other based on the coverage area. The major applicable area is when a source UAVs are required to search the assigned target area and send the updates to the designated Ground base station with the help of intermediate nodes.


\subsubsection{\textbf{Assumptions used in the proposed RLPR}}
\begin{enumerate}[i]
	\item All nodes in the network are equipped with IEEE 802.11 \cite{marconato2016ieee} wireless devices and operate at 2.4 GHz \cite{fabra2017impact} of frequency and each node is equipped with TCP/IP architecture. 
	\item	Each and every node has a unique identifier (ID) to distinguish itself with other UAVs in the network
	\item	Each node in the network is equipped with a GPS device to acquire its location. 
	\item There is no disjoint path in the network.
	\item All nodes are moving with random speed and random-way point mobility model is used, where maximum speed of nodes in the network is 25km/h.
	\item Nano UAVs are used that are flying at altitude of 100m and have transmission range less than or equal to 250 m.
	\item Free space radio propagation model is used for simulation with simulation area of $1000 \times 1000\mathop m\nolimits^2 $.
	\item Destination is fixed and geographic position of the destination is known to all the nodes.
	\item Energy threshold of Network is 10 Joule and RSSI~threshold of network is -64 dBm.	
\end{enumerate}

\subsubsection{\textbf{Forwarding zones criterion}}
The forwarding zone is established at each energy-efficient and reliable node in the network towards the destination for a certain data flow. One of the criteria in selecting the forwarding zone is to select that particular node that has a longer lifetime in terms of energy and node is moving toward the destination and shall maintain the connection for a longer duration within a limited region according to the maximum defined transmission range. Also, all the other nodes are rejected to form a forwarding zone. One of the advantages of establishing a forwarding zone is that it decreases the possibility of selecting those nodes which are far away from the destination based on the vicinity. This usually decreases the ETE delay in the network in a single source and destination scenario. For this purpose, every node sets a forwarding zone and only those nodes are allowed to rebroadcast the message which lies inside the forwarding zone of a particular node. The procedure of selecting the next hop in the forwarding zone is as follows.

\begin{enumerate}[i]
	
\item The node that wants to select the next-hop node from the forwarding zone first checks whether its distance to destination is within the defined communication range, $R$. If the destination falls within the defined threshold of a forwarder, then it means that the destination is within the communication range of forwarder and it can directly send as a unicast packet to the destination. In contrary to this, the forwarder embeds the forwarding angle in the packet and broadcasts the packet to its 1-hop neighbors.
\item Receiving node computes its energy if node energy is greater than some defined threshold level then receiving node computes forwarding angle and checks whether it is in forwarding zone of previous relay node based on information in packet receive from previous forwarding node. If it is not in the forwarding zone, it discards the packet. Forwarding zone is determined through the forwarding angle 	 $\theta$ and the maximum value of forwarding angle is used as ${180^ \circ }$.Forwarding angle is computed using  Equation~\ref{E1} as also discussed in \cite{li2015self}. 
	
	\begin{equation}
	\label{E1}
	\theta = {\cos ^{ - 1}}\left| {\frac{{_{\mathop x\nolimits_{rn} - \mathop {\mathop x\nolimits_{ph} }\nolimits_{} }}}{{\sqrt {(\mathop {\mathop x\nolimits_{rn} - \mathop x\nolimits_{ph} )}\nolimits^2 - (\mathop {\mathop x\nolimits_{rn} - \mathop x\nolimits_{ph} )}\nolimits^2 } }}} \right| \le \mathop {180}\nolimits^ \circ 
	\end{equation}  
\\
	Consider the network with 12 nodes as depicted in Figure~\ref{fzEX}. Here, the angle to compute the forwarding zone is set as ${180^ \circ }$. All receiving nodes based on the defined vicinity as per the maximum transmission range calculates their energy and if their energy is greater than defined threshold nodes computes forwarding angle. If the value of the forwarding angle is less than ${180^ \circ }$ then the receiving nodes are considered to be in the forwarding zone and are allowed to rebroadcast the received packet. All the other nodes that do not satisfy the condition, shall discard the received packet. Node 1 is the source and node 12 is the destination node, and destination node is not in the communication range of the source node. Node 1 broadcast the message in the network upon receiving the message each node first computes their energy if their energy is greater than define threshold level then nodes computes forwarding angle with respect to node 1. Node 6 and 7 lies in the forwarding zone of node 1 whereas node 5 energy is less than some defined threshold level so due to that it does not lie in the forwarding zone of node 1. Now nodes 6 and 7 rebroadcast the packet in the network, whereas nodes 2, 3, 4, and 5 do not satisfy the forwarding zone condition and discard the packet.

	\subsubsection{\textbf{Front relative nodes}}
All the nodes which lie in the forwarding zone of a particular node within 1-hop neighbors are nominated as the front relative nodes because of the forwarding zone. Additionally, all such nodes have the direction towards the destination. This selection mechanism can decrease the control overhead message in the network. Now when a source node needs to discover a route, it broadcasts a message in the network; only those nodes are allowed to rebroadcast that are towards the destination and must be the part of the front relative list of the predecessor node. In this way highly connected network among 1-Hop neighbors is achieved and that highly connected chain starts from the source node and terminated at the destination node. From Figure~\ref{fzEX} it is shown that front relatives of node 1 are node 6 and node 7.

	\begin{figure}[pos=t]
		\centering
		\includegraphics[scale=0.59]{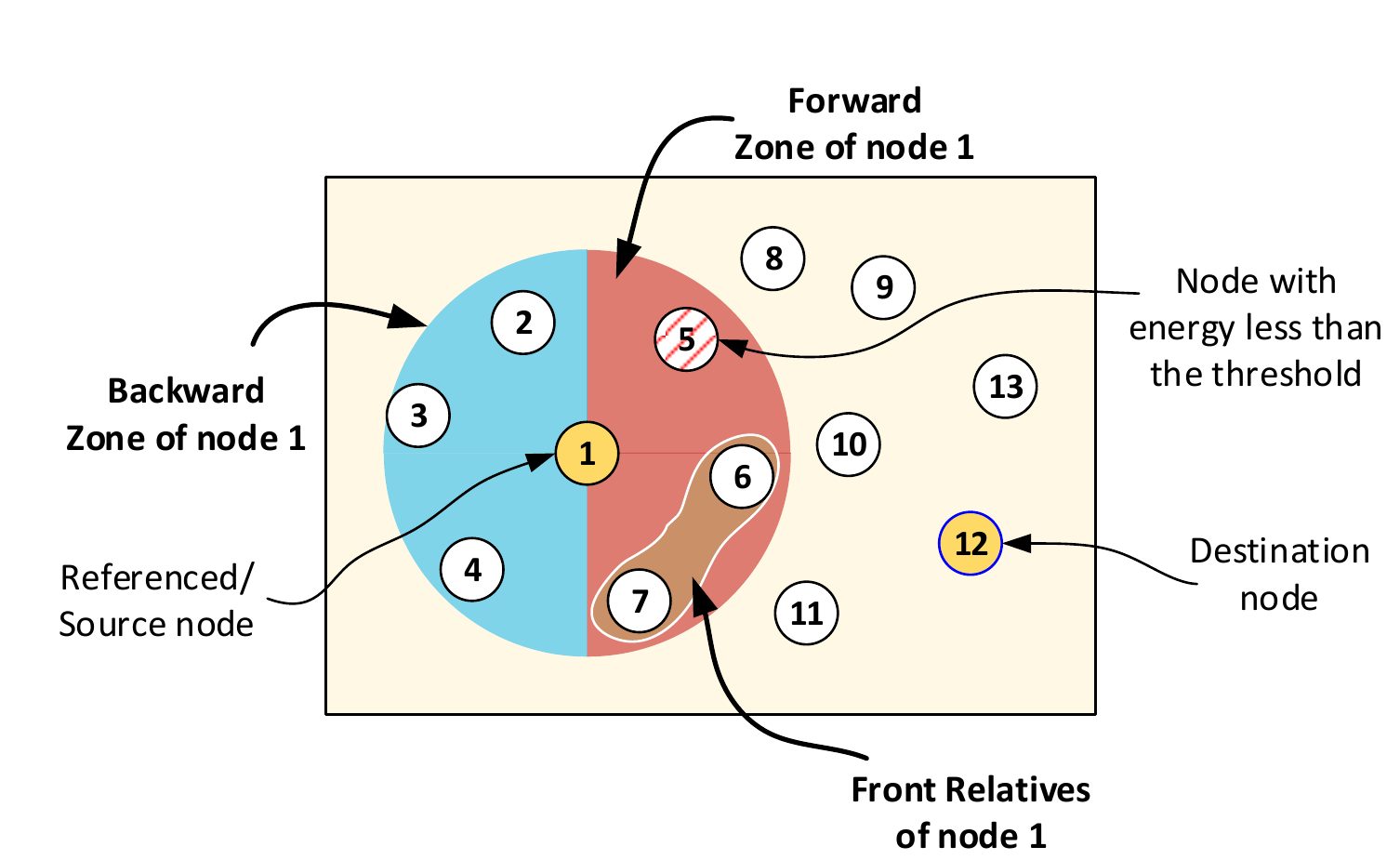}
		\caption{Illustration of Forwarding Zone computation in RLPR}
		\label{fzEX}
	\end{figure}

\end{enumerate}
\subsection{\textbf{Routing metric for RLPR}}
In this section, certain routing metrics for selecting the routing path that can enhance the network performance is discussed in detail.

\subsubsection{\textbf{Received signal strength}}
High Received signal strength (RSS) is considered as an indication for better link quality and the distance between two nodes. To achieve high connectivity level between the nodes, every node upon reception of a packet calculates its signal strength; if it is greater than the defined threshold value, then it rebroadcasts the packet in the network, otherwise, it discards the packet.

\subsubsection{\textbf{Geographic distance of nodes towards destination}}

As already discussed, that if a node fulfills the forwarding zone criteria, then it is allowed to rebroadcast the packet in the network. Another way to enhance the network performance is to select that relay or next-hop node from the forwarding zone whose geographical progress is towards the destination and should have to be within the communication range for a longer duration. By computing this metric, a node that is closer to the destination and that has high link stability is selected as a next forwarding node in the network. This procedure will be carried out until the destination is reached. Now each and every front relative node in the network computes the aforementioned metric. 

The (GD)~\cite{rezende2012virtus} of a node towards the destination is computed using the following Equation~\ref{E2}.

 \begin{equation}
\label{E2}
GD = \left| {1 - \frac{{\mathop {\left| {DP} \right|}\limits^ \to  - \mathop {\left| {DN} \right|}\limits^ \to }}{r}} \right|
\end{equation}
$\mathop {\left| {DP} \right|}\limits^ \to $is distance from previous forwarding node to destination and
$\mathop {\left| {DN} \right|}\limits^ \to $is distance from receiving node to destination and $r$ is maximum transmission range of node.

\subsubsection{\textbf{\textbf{Relative speed of a node}}}

To achieve a high connectivity level among the nodes, the relative speed is computed. When a node receives any packet from its previous node, it calculates relative speed with respect to the previous relay node. Node with analogous relative speed rebroadcasts the packet in the network. The relative speed between the nodes is computed through the following Equation~\ref{E3}.

\begin{equation}
\label{E3}
\mathop V\nolimits_{RL} = \left| {\frac{{\mathop V\nolimits_{ph} - \mathop V\nolimits_{rn} }}{{\mathop V\nolimits_{\max } }}} \right| 
\end{equation}
\\
$\mathop V\nolimits_{ph}$ is speed of previous forwarding node, $\mathop V\nolimits_{rn}$ Speed of receiving node where as $\mathop V\nolimits_{max}$ is maximum speed of node in the network

\begin{figure}[pos=t]
	\centering
	\includegraphics[scale=0.6]{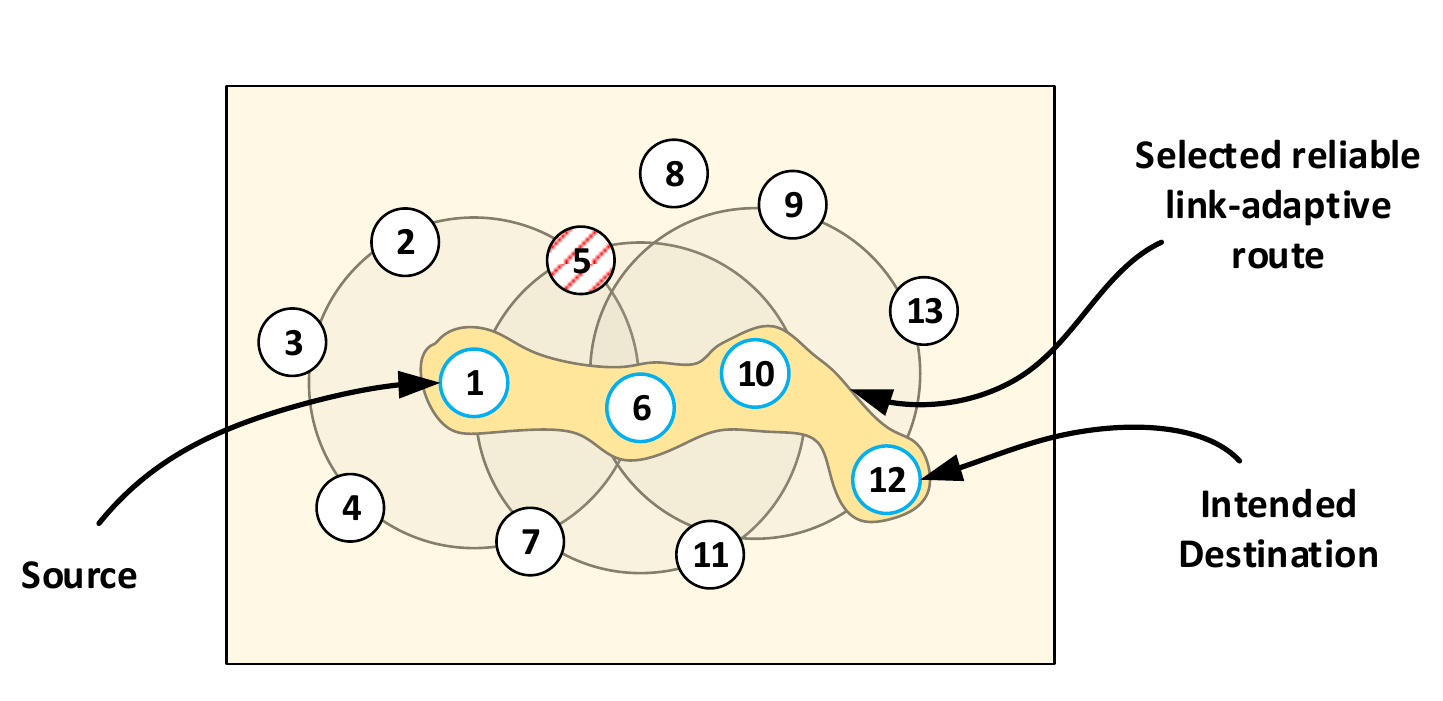}
	\caption{Network Scenario to Determine the Composite Routing Metric}
	\label{cmex}
\end{figure}

\subsubsection{\textbf{\textbf{Composite routing metric }}}
The composite routing metric is computed from several routing metrics as defined in Equations~\ref{E2} and~\ref{E3}. By using a composite metric, a node with high connectivity level and whose position toward the destination is selected as a next forwarding node. Each node in the network computes the composite metric set its timer and node whose timer expires first will rebroadcast the packet in the network and all other nodes in the network will discard the packet and reset their timer. Composite metric is calculated by using the following Equation~\ref{E4}:
\begin{equation}
\label{E4}
{\rm{Composite\;metric}} = \alpha \times GD + \beta \times \mathop V\nolimits_{RL} 
\end{equation}
here~$\alpha$ and $\beta$~are the weights with equal weightage. 

With reference to Figure~\ref{cmex}, Node 1 is a source node and its front relatives are Node 6 and Node 7. Node 12 is a destination node and it is not in the vicinity of Node 1. All the front relative nodes upon receiving a packet from Node 1 calculates the composite metric using  Equation~\ref{E4}  and set a timer. Both front relative Node 6 and Node 7 compute composite metrics and start their timer. The Node whose timer expires first among all the nodes in the neighbourhood will forward the route request in the network. 
With reference to the computed values, Node 6 timer will expire first and it will rebroadcast the message to all the other nodes within its vicinity. Node 7 will discard the route request and reset its timer. It is to be noted that the geographic distance of Node 6 and Node 7 is 0.44m and 0.72m, respectively. Similarly, the relative speed considered to select the appropriate node in the above particular scenario for Node 6 and Node 7 is 0.08m/s and 0.16m/s, respectively. Therefore, the values of composite metric that has been computed for Node 6 and Node 7 are 0.26 and 0.44, respectively.

\subsection{\textbf{Working principle of RLPR}}
This section briefly describes the working principle of the proposed routing protocol that is RLPR. Three different types of messages are exchanged between the nodes in a network to achieve the desired objectives. The periodic 1-hop HELLO broadcast message along with an adaptive zoom out interval. HELLO-message is a periodic broadcast message in the network that is used to show the presence and to share the geographical position, speed, distance, and energy of nodes in the network. In contrast to this, the zoom-out HELLO message is triggered, if any change at the link-layer level of the TCP/IP layering model is detected. A Reliable link route request (RLRQ) message is a broadcast message that is broadcast by the source node in the network. When a source node needs to send the data to the destination and a reliable and energy-efficient path is not available for the required destination, it generates RLRQ. When an RLRQ message reaches the destination, it creates an RLRP (Reliable link route reply)  message which is a unicast message sent by the destination.

\subsection{\textbf{1-Hop/Zoom-Out broadcast messages} } 
\begin{figure}[pos=b]
	\centering
	\includegraphics[width=0.8\linewidth]{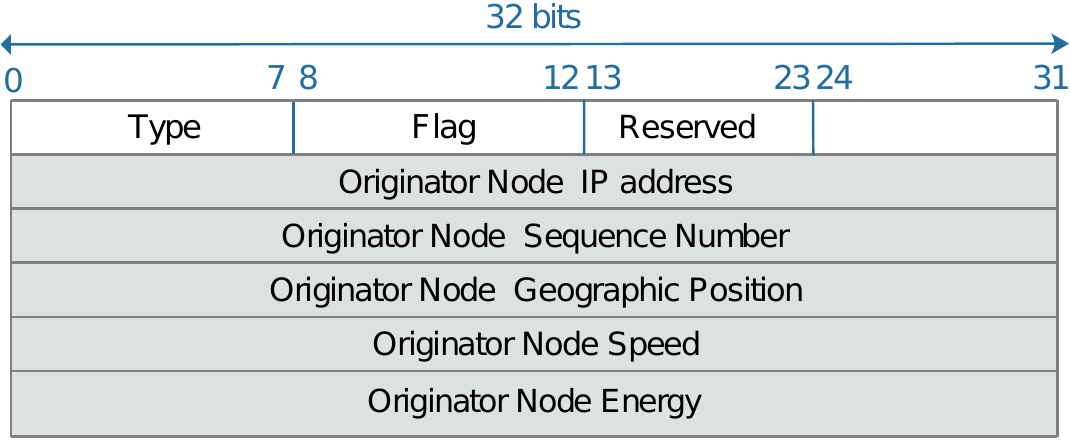}
	\caption{1-Hop Hello Broadcast Message Format of RLPR}
	\label{hello2}
\end{figure}

Two types of 1-hop broadcast messages are used in RLPR. The first one is the 1-Hop HELLO broadcast message that is generated by each and every node in the network, periodically. The periodic interval for 1-Hop HELLO message is used as 1sec. Along with this, an adaptive zoom out 1-Hop message broadcasted when a change in the network is detected. The Hello message format is shown in Figure~\ref{hello2}.

Steps involved in 1-Hop HELLO broadcast message are as follows: 
\begin{enumerate}[i]
\item  Each node in the network sends Hello message with its 1-Hop Neighbors if its energy is greater then threshold level. Figure \ref{hello2} shows the information that is exchanged through Hello broadcast packet.
\item Based on the information from 1-hop Hello message, each node calculates, and stores the necessary information of its 1-hop neighbor in the neighbor table. All receiving nodes based on the defined vicinity as per the maximum transmission range calculates their energy. Nodes whose energy levels are greater than the defined threshold energy level play the role for further processing. 
\item	The node which fulfills the criteria mention in step 2 checks whether its 1-Hop neighbors lie in the forwarding zone by calculating the forwarding angle. If a 1-hop neighbor lies in the forwarding zone, then the node receives the information and store it in the front relative neighbor table. 

\end{enumerate}

\subsection	{\textbf{Reliable link route request (RLRQ)} }
\begin{figure}[pos=t]
	\centering
	\includegraphics[width=0.9\linewidth]{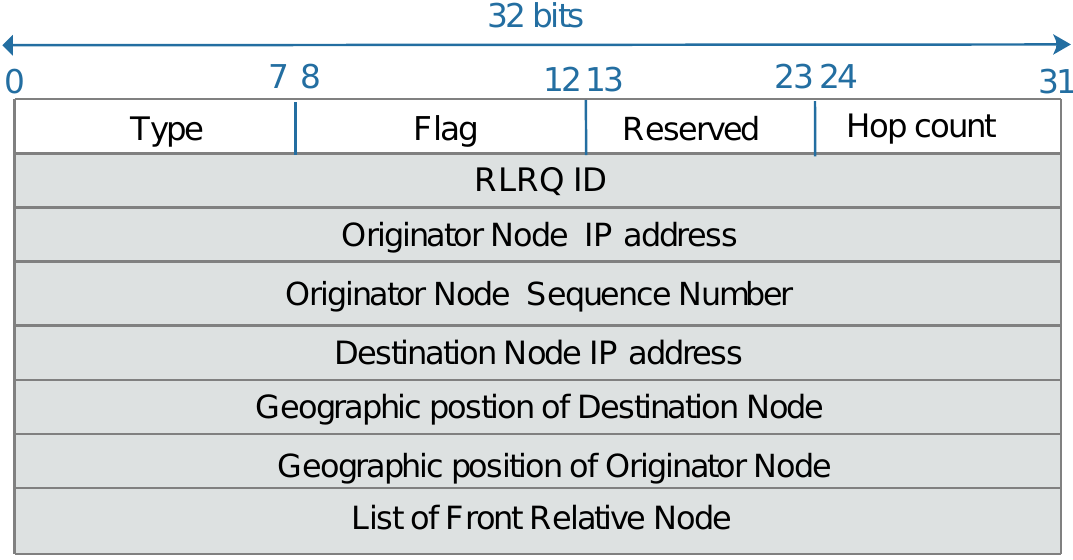}
	\caption{RRLQ Message Format}
	\label{rlrq1}
\end{figure}

The following steps show the description of RLRQ message.
\begin{enumerate}[i]
\item RLRQ message is broadcast to all its 1-Hop neighbors in the network. When a source node wants to send the data to the destination and source node does not have an optimal path it broadcast RLRQ. A node that receives the RLRQ message is a destination or it has a path to the destination, it will unicast a route reply message. RLRQ packet shares the information as shown in Figure~\ref{rlrq1}.
	
\item If the node is not a destination or it has no optimal path to destination then the node will rebroadcast the RLRQ message in the network. Only those nodes will rebroadcast the message that is a front relative of the previous forwarding node, otherwise, they will discard the packet. 
\item If the front relative receives the RLRQ message, it will compute the received signal strength and the front relative that signal strength then is lower than the defined threshold level will discard the packet. 
\item Node whose Signal strength is greater than the defined threshold level plays the role for further processing. Now front relative nodes that signal strength is greater than define threshold level calculate relative speed with respect to the previous forwarding node.
\item On receiving the RLRQ message, the nodes determine the geographical distance with respect to the destination. 
\item Each and every front relative node calculates the composite metric and sets a waiting time. The node whose waiting time expires first to rebroadcast the RLRQ message in the network. When the nodes who have stared their timer receive an RLRQ packet for the same path based on the source, destination, broadcast, and sequence number identifiers, it discards the packet, otherwise it rebroadcast the RLRQ message when its timer expires. 
\item When the RLRQ packet is rebroadcasted by an intermediate node, only the front relative will able to accept the packet, all the others node will discard the packet as per the criterion defined in the forwarding zone. All the front relatives will discard the packet if they already have received the RLRQ packet for the same source and the same destination with the same sequence number, however, the next step will be carried out from steps 2 to 6. If the RLRQ packet reaches the destination, it will unicast Reliable link route reply (RLRP) packet.

\end{enumerate}

\begin{figure}[pos=hbt]
	\centering
	\includegraphics[width=1.0\linewidth]{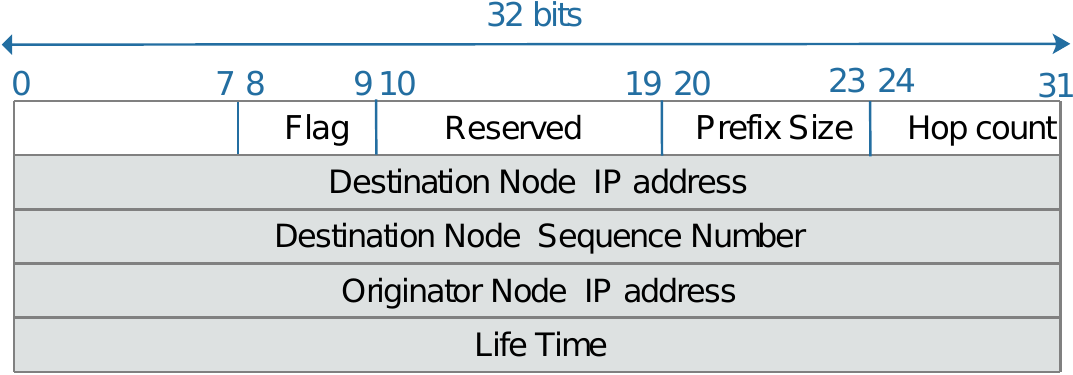}
	\caption{RLRP Message Format}
	\label{fig:replyformat}
\end{figure}

\subsection{\textbf{Reliable link route reply (RLRP)}}
When the RLRQ message arrives at the destination, it maintains a backward table and sends a unicast route reply message along a reliable path based on the backward entries maintained during RLRQ. The RLRP message format is shown in Figure~\ref{fig:replyformat}. If an intermediate node receives an RLRP message, it will maintain a forward table and resend a Unicast RLRP message to the next hop until the packet arrives at the source node. When the source node gets the RLRP message, it will start data forwarding.

\begin{table} [pos=b]
	\centering
	\caption{Notations Used in RLPR Algorithms}
	\label{tab:my-table6}
	\begin{tabular}{l|l} 
		\hline
		\textbf{Notation} & \textbf{Description}               \\ 
		\hline
		\textit{N}                                       		& Number of Nodes in the Network          \\ 
		\hline
		$\mathop N_i$            												& $i^{th}$~Node in the Network           \\ 
		\hline
		
		$\mathop P_i~ $            & $\{ x,y,z\} $ position~ coordinates of node $i$ \\ 
		\hline
		$\mathop Z_i $           &Forward route entries of node $i$                \\ 
		\hline
		$\mathop \rho _i$  & Reverse route entries of node $i$~                \\ 
		\hline
		$\mathop V_i $           & Speed of Node $i$                                \\ 
		\hline
		$\mathop A_i $           & Neighbors of node $i$                              \\ 
		\hline
		$ E_i $                          & Energy of Node $i$                               \\ 
		\hline
		$\mathop T_E $           & Energy Threshold                            \\ 
		\hline
		$\mathop {S}_i $         & Signal strength of Node $i$                           \\ 
		\hline
		$\mathop T_ {s} $         & Signal Strength Threshold                             \\ 
		\hline
		$\mathop {FR}_i$         & Front Relative Table of Node $i$                        \\ 
		\hline
		$\mathop {R}_i$          & Neighbor Table of Node $i$                           \\ 
		\hline
		$\mathop \theta _i $ & Forwarding angle of Node $i$                          \\ 
		\hline
		$\mathop S\nolimits_{id}$         & Source Node id                                  \\ 
		\hline
		$\mathop D\nolimits_{id}$         & Destination Node identification                                  \\
		\hline
		
		$\mathop B\nolimits_{id} $         & Broadcast identification      \\ 
		\hline
		$\mathop {Vr}_{i} $         & Relative speed of Node $i$                           \\ 
		\hline
		$\mathop {Gd}_i $         & Geographic Distance of Node $i$                         \\ 
		\hline
		$\mathop \tau _i $  & Timer for Node $i$                               \\
		\hline
	\end{tabular}
\end{table}


\subsection{\textbf{Algorithm for RLPR}}
The process of discovering the route from source to destination using forwarding angle and the front relatives based on the defined routing criteria is elaborated in accordance with Algorithm-\ref{algo_1}, Algorithm-\ref{algo_2}, and Algorithm-\ref{algo_3}. The notation used in these algorithms is tabulated in Table~\ref{tab:my-table6}. Each and every node in the network is required to share the geographical position and energy to its 1-hop neighbors using Algorithm-\ref{algo_1}. Additionally, each and every receiving node computes its forwarding angle and selects its front relative and populate/update front Relative table. Furthermore, Algorithm-\ref{algo_2} is used, when a source node wants to send a data to an intended destination. Moreover, if a source node does not have any path to destination, it broadcasts RLRQ packet. This packet contains the Source ID, Destination ID, list of front relative nodes, Speed, and Position. If receiving node is in the front relative nodes, it computes routing metric for path selection and starts its timer. When a destination node receives RLRQ packet, it unicasts the Route reply packet using Algorithm-\ref{algo_3}. If receiving nodes is a source node, it forwards data packet and the rest of the nodes with valid entries, update their forward tabled and send unicast RLRP packet.

\begin{algorithm}[b]
\label{algo_1}
	\SetAlgoLined
	
	\textbf{Input}: $N$, $P_i$, $V_i$, $E_i$, $\theta_i$, $B_{id}$, $A_i$  \\
	\textbf{Output} : $R_i$, $F_{r_i}$
	
	Initialization\;

\DontPrintSemicolon

	\For{Each Node $i \in N$}{
		\tcc{Assign position cordinates to each node and Update Neigbour Table } 
$\mathop R\nolimits_i \leftarrow \left\{ {\mathop A\nolimits_i ,\mathop E\nolimits_i ,\mathop P\nolimits_i } \right\}$

		\If{$\mathop E\nolimits_i \ge \mathop T\nolimits_E $}{
	Broadcast 1-Hop Hello Broadcast Message 
}
}
	\For{Each Node $i \in N$}{	\tcc{For each node $i$ that receives the broadcast message}
		\uIf{$(\mathop i \in \mathop R\nolimits_i )~\& \&~ (\mathop E\nolimits_i \ge \mathop T\nolimits_E )$}{
			\tcc{For All Receiving Node $\mathop \forall \nolimits_i\in N$, Compute Forwarding angle}  
Compute~$\mathop \theta \nolimits_i $ using Equation \ref{E1}
			\\ \If{$\mathop \theta \nolimits_i \le 180^ \circ$} {
				\tcc{ Populate and Update Front Relative Table}
				$\mathop {Fr}\nolimits_i \leftarrow \mathop R\nolimits_i    $         $\mathop {\therefore A}\nolimits_i \in \mathop R\nolimits_i $
				$ $
			}
	}
		\Else{
			Discard the Entry			
		}
	}
	
\caption{\textbf{1-Hop/Zoom-out Hello Broadcast Message}}
\end{algorithm}

\begin{algorithm}[t]
\label{algo_2}
	\SetAlgoLined
	\textbf{Input}: $S_{id}$, $D_{id}$ , $\theta_i$, $F_{r_i}$, $P_i$, $V_i$, $B_{id}$  \\
	\textbf{Output} : $Z_i$, $\rho_i$
	
	Initialization\;	
	$\mathop S\nolimits_{id} = Null,\mathop D\nolimits_{id} = Null,\mathop P\nolimits_{i} = \mathop B\nolimits_{id} = \mathop {Fr}\nolimits_i = \mathop V\nolimits_i = 0$\\
	Request to send Data packet
	
	\uIf{$ S_{id}$ has path to $D_{id}$}
	
		\indent~~~Forward data packets using information in $\mathop \rho \nolimits_i$
\\	\Else{
		
		Broadcast RLRQ packet  
		$\mathop S\nolimits_{id} ,\mathop D\nolimits_{id} ,\mathop P\nolimits_{i}, \mathop B\nolimits_{id} ,\mathop {Fr}\nolimits_i ,\mathop V\nolimits_i $	
		
		\While{($ i ! = \mathop D\nolimits_{id} $) $\mathop \forall \nolimits_i \in N$}         
		{
			
			\uIf{Reiceving Node $i$ has a path to $\mathop D\nolimits_{id} $}{
				\tcc{ Reiceving Node is Destination Node or it has Path to Destination}
				Go to step 23
			}
		
	\uElseIf{$ i \in \mathop {Fr}\nolimits_i $}{
	\tcc{ if Recieving node is Front Relative Node Then Compute Composite Metric}
	Compute $\mathop {S}\nolimits_i $
	\\	\If{$\mathop {S}\nolimits_i $$>$=$\mathop T\nolimits_{s}$}
	{
		Compute
		$\mathop V\nolimits_{Ri}$, $\mathop {GD}\nolimits_i$ using Eq-\ref{E2} and \ref{E3}
		\\Compute Composite Metric using Equation \ref{E4} and start time $\mathop \tau \nolimits_i $ 
		
	}
	\If{$\mathop \tau \nolimits_i $ Expire}
	{
		
		Update $\mathop \rho \nolimits_i $
		Rebroadcast RLRQ message
	}
}
			\Else{
				Discard Packet
			}
		\If{$i = = \mathop D\nolimits_{id} $} 
			{
				\tcc{Destination node recieves RLRQ packet }
				Update $\mathop Z\nolimits_i $ 
				\\	Unicast Reliable Link Route Reply
			}	
		}
	}
\caption{\textbf{Reliable Link Route Request Procedure of RLPR}}
\end{algorithm}

\begin{algorithm}[hbt]
	\label{algo_3}
	\SetAlgoLined

\textbf{Input}: $S_{id}$, $D_{id}$ , $B_{id}$  \\
	\textbf{Output} : $Z_i$
	
	Initialization\;
	
	\DontPrintSemicolon
	
	\For{Each Node $i \in N$}{
		\uIf{$i =\mathop S\nolimits_{id} $}
		{
			\indent Forward Data Packets on the available Path
		}
		\Else{
			\tcc{Update Forward Route Entry }
			Update $\mathop Z\nolimits_i$
			\\	Unicast RLRP
			GO to step 3	
		}
	}
\caption{\textbf{Reliable Link Route Reply Procedure of RLPR}}
\end{algorithm}

\subsection{\textbf{Performance evaluation}}
  For evaluation of the RLPR, we used the following performance evaluation parameters for the proposed PLPR. 
  \subsubsection{\textbf{Control packet overhead}} 
Routing protocol needs to send control information in order to find a routing path. It is important to examine that how much control information is sent by each protocol in the network. This is because, whenever a node needs to send the information, it needs to acquire the channel. If the channel acquisition rate is high, then the chances of collision increase. This increases the energy consumption that leads to a decrease the lifetime of the network. In the RLPR protocol, a number of relay nodes in the network is less, this reduces control messages overhead. As only those nodes that are in the forwarding zone of the Previous-hop will relay the route request and all others will discard the packet.  
  \subsubsection{\textbf{Search success rate}}
The search success rate can define as the ratio of a total number of route request messages and route reply messages that are broadcast by nodes for path discovery to total time taken by the network to discover the path. 
  
 \subsubsection{\textbf{  Network lifetime}}
Network lifetime can be defined as the time from network initialization to first node failure because of battery depletion. Failure of a node may result in forming a disjoint path in the network. In a network, if an intermediate node is bottleneck node it decreases the network lifetime as the energy of this bottleneck node is consume more quickly and this cause disjoint path in the network Reliability of the Network can be achieved if Network has the better Network lifetime. In our proposed mechanism Reliability of the network is achieved as that next relay node in the forwarding zone is selected with a better energy level.
\section{\textbf{Simulation and results} }
\label{results}

This section presents simulation results and performance analysis of the proposed RLPR routing protocol. RLPR is implemented using Network simulator (NS-2.35). Each simulation scenario ran for 30 times with random seed value, to get 95\% confidence so that the results show an admissible range along with average, minimum, and maximum values. For the simulation analysis, it is assumed that there is only a single destination in the network for all the scenarios; however, with a varying number of sources. Additionally, a random waypoint mobility model is used, by considering a random speed between a range of 10 - 25 km/h.

Furthermore, a free space radio propagation model is used for simulation and each node in the network is equipped with an omnidirectional antenna that has the capability of transmitting and receiving to/from all directions, respectively. Moreover, Constant bit rate (CBR) traffic is used over the User datagram protocol (UDP) of the transport layer with a packet size of 512 bytes. Simulations are carried out for three different routing protocols; AODV, RARP, and RLPR. Whereas, the performance evaluation metrics simulated for each protocol are (1) control message overhead, (2) search success rate and (3) network lifetime. The proposed RLPR protocol is compared with RARP and with the conventional AODV and by varying the numbers the nodes, a number of sources, and simulation time.

\begin{table}[pos=t]
	\centering
	\caption{Simulation Parameters}
	\label{tab:my-table}
	\begin{tabular}{|c|c|}
		\hline
		\multicolumn{1}{|l|}{\textbf{Physical Layer Parameters}}          & \textbf{Values}       \\ \hline
		\begin{tabular}[c]{@{}c@{}}Radio Propagation\\ Model(outdoor)\end{tabular} & Free space         \\ \hline
		Mobility Model                               & Random way point mobility  \\ \hline
		Reception threshold                            & -64dbm           \\ \hline
		Antenna model                                & Omni-directional      \\ \hline
		Speed of Nodes                               & 10-25 km/h         \\ \hline
		Carrier Frequency                              & 2.4Ghz           \\ \hline
		Maximum transmission Range                         & 250m            \\ \hline
		Energy threshold                              & 10 Joules          \\ \hline
		\multicolumn{2}{|l|}{\textbf{MAC layer}}																 \\ \hline
		CW min                                   & 15             \\ \hline
		CW max                                   & 1023            \\ \hline
		Slot time                                  & 20             \\ \hline
		\multicolumn{2}{|l|}{\textbf{Network layer}}																 \\ \hline
		Routing Protocol                              & RLPR, AODV, RARP       \\ \hline
		\multicolumn{2}{|l|}{\textbf{Transport layer}}																 \\ \hline		
		Protocol                                  & UDP             \\ \hline
		Interface queue type                            & Queue/DropTail/priQueue 	 \\ \hline
		Interface queue length                           & 10             \\ \hline
		\multicolumn{2}{|l|}{\textbf{Application layer}}																 \\ \hline		
		Application                                 & CBR             \\ \hline
		Packet size                                 & 512 Bytes          \\ \hline
	\end{tabular}
\end{table}

\begin{figure}[pos=t]
	\centering
	\includegraphics[scale=0.26]{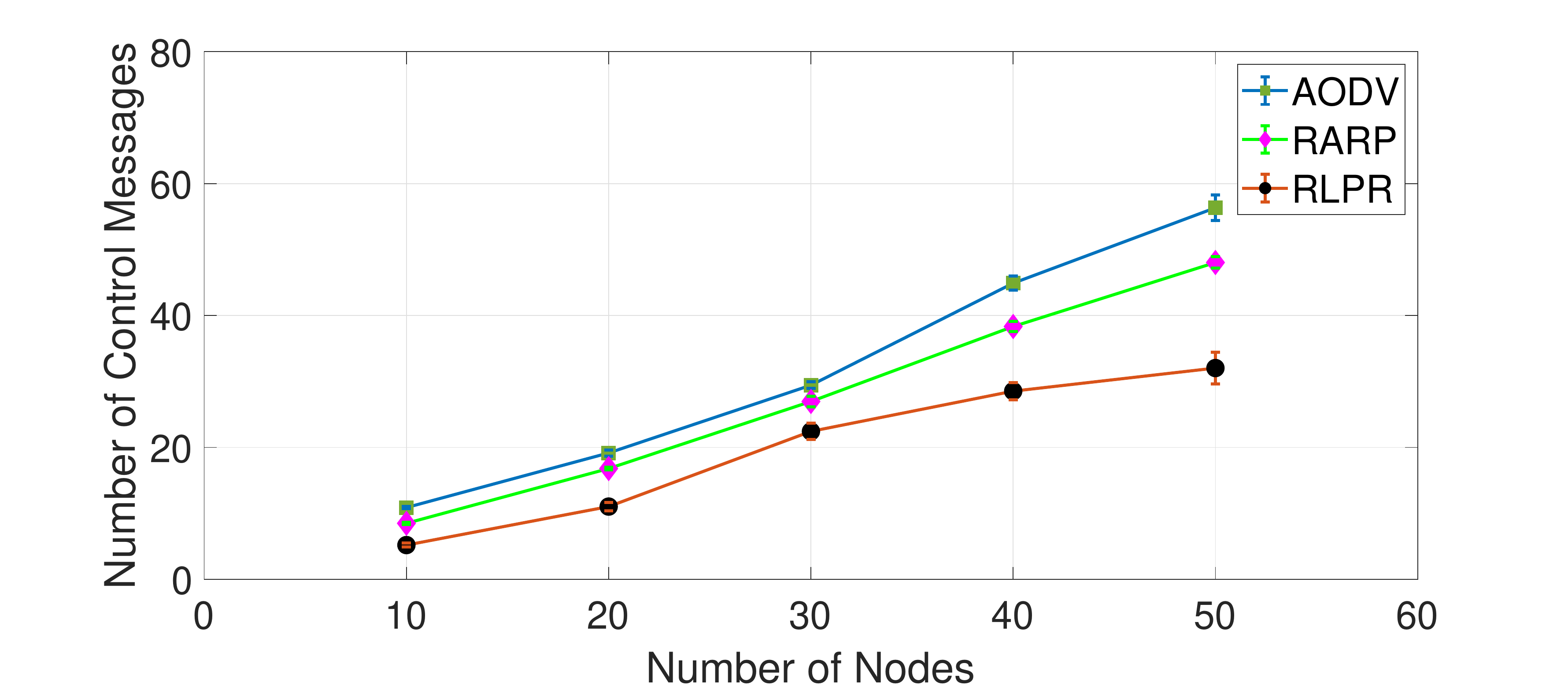}
	\caption{Impact of Number of Nodes on Control Messages}
	\label{noofnodecm}
\end{figure}

Figure~\ref{noofnodecm} shows the control message overhead generated in the network by varying the number of nodes in a single source and destination scenario. The horizontal axis represents the numbers of nodes and the vertical axis shows the corresponding control messages generated by the nodes in the network. It has been observed that the control message overhead increases with the increase in the number of nodes in the network using RLPR, RARP, and AODV routing protocols. However, RLPR has shown better results as compared to RARP and AODV in terms of control message overhead. This is because, in RLPR, only front relative nodes, which are in the forwarding zone and are towards the destination are allowed to rebroadcast the message in the network. The nodes which are not in the forwarding zone, simply discard the message.

\noindent 

\begin{figure}[pos=b]
	\centering
	\includegraphics[scale=0.264]{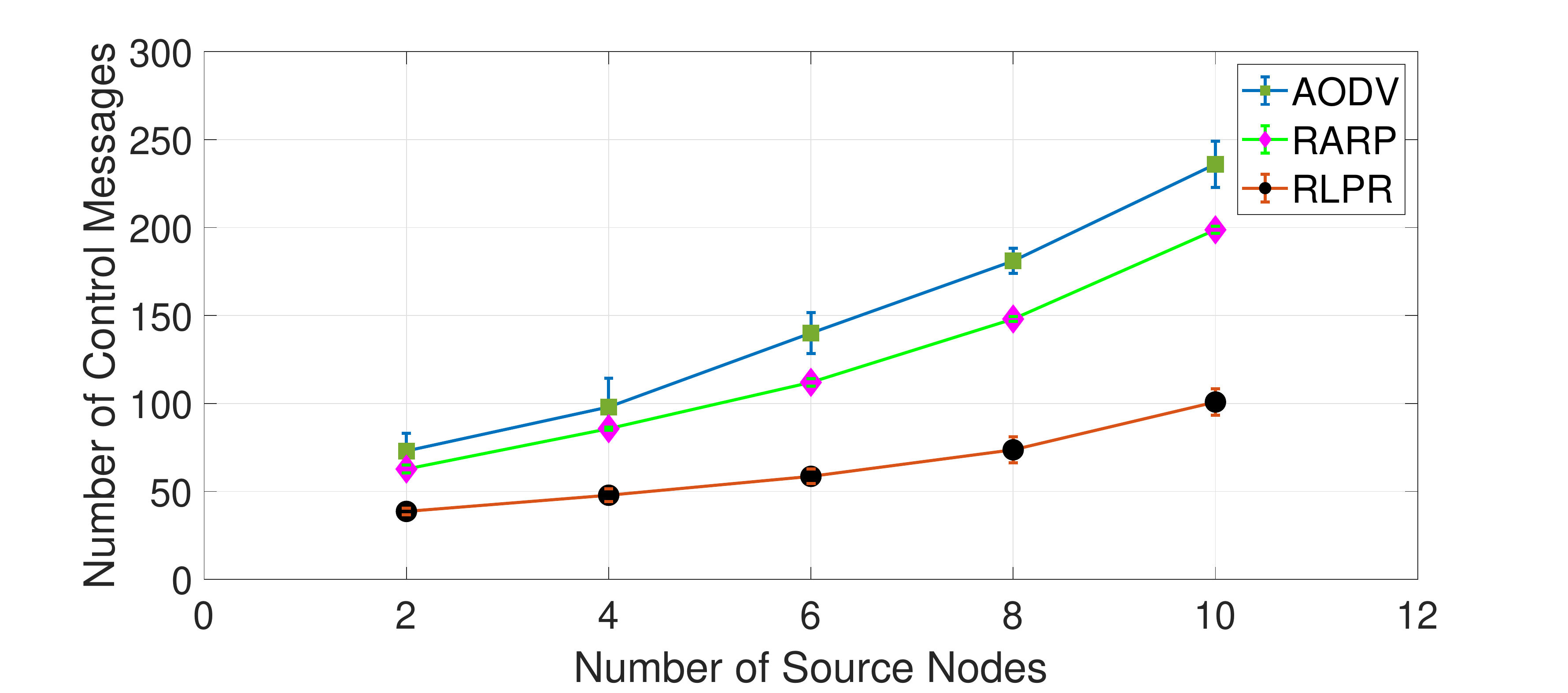}
	\caption{Impact of Number of Source Nodes on Control Messages}
	\label{noofsrscm}
\end{figure}

Additionally, in RLPR, nodes are selected on the basis of received signal strength and speed of the nodes; this leads to an increase in the reliability in terms of connectivity level that ultimately decreases the control message overhead as compared to RARP and AODV. In contrary to this, RARP and AODV allow all the nodes in the network to re-broadcast the request/message. However, RARP shows better results as compared to AODV. This is because, RARP uses the Omni-directional antenna while broadcasting the request and directional transmission towards the predicted location using a directional antenna, that enables a longer transmission range, and therefore, the control message overhead decreases. On the other hand, AODV generates higher control message overhead as compared to RARP and RLPR. This is due to the fact that AODV just uses the hop count criterion to discover the route between source and destination, so there might be a chance that the nodes that are selected in a route may become a bottleneck, this increases the packet loss that leads to rebroadcast the message for better ETE route.

Figure~\ref{noofsrscm} shows the control message overhead generated in the network by varying the number of sources. In this scenario, a single destination with 30 numbers of nodes in the network is considered. The horizontal axis represents the number of sources and the vertical axis shows the corresponding control messages generated by the nodes in the network. It has been observed that the control message overhead increases with the increase in the number of sources in the network using RLPR, RARP, and AODV routing protocols. Using RLPR significantly decrease the number of control message as compared to AODV and RARP. This is because in RLPR each and every source node initially computes its own front relative nodes and only front relative nodes of every source node are allowed to re-broadcast the route request in the network. The rest of the nodes which are in the backward zone do not rebroadcast the packet. On the other side, RARP and AODV allow all the nodes in the network to rebroadcast the messages/route-requests. However, RARP shows better results as compared to AODV. This is due to the usage of a utility function to discover the route. The utility function is computed on the basis of longer connection time between nodes and the minimum risk value, where risk value is computed on the basis of nodes' energy and their geographical position, such as environmental conditions and terrain structure. On the other hand, AODV just uses the hop count criterion to discover the route between source and destination. It allows all the nodes to rebroadcast the message once.

\begin{figure}[pos=b]
	\centering
	\includegraphics[scale=0.26]{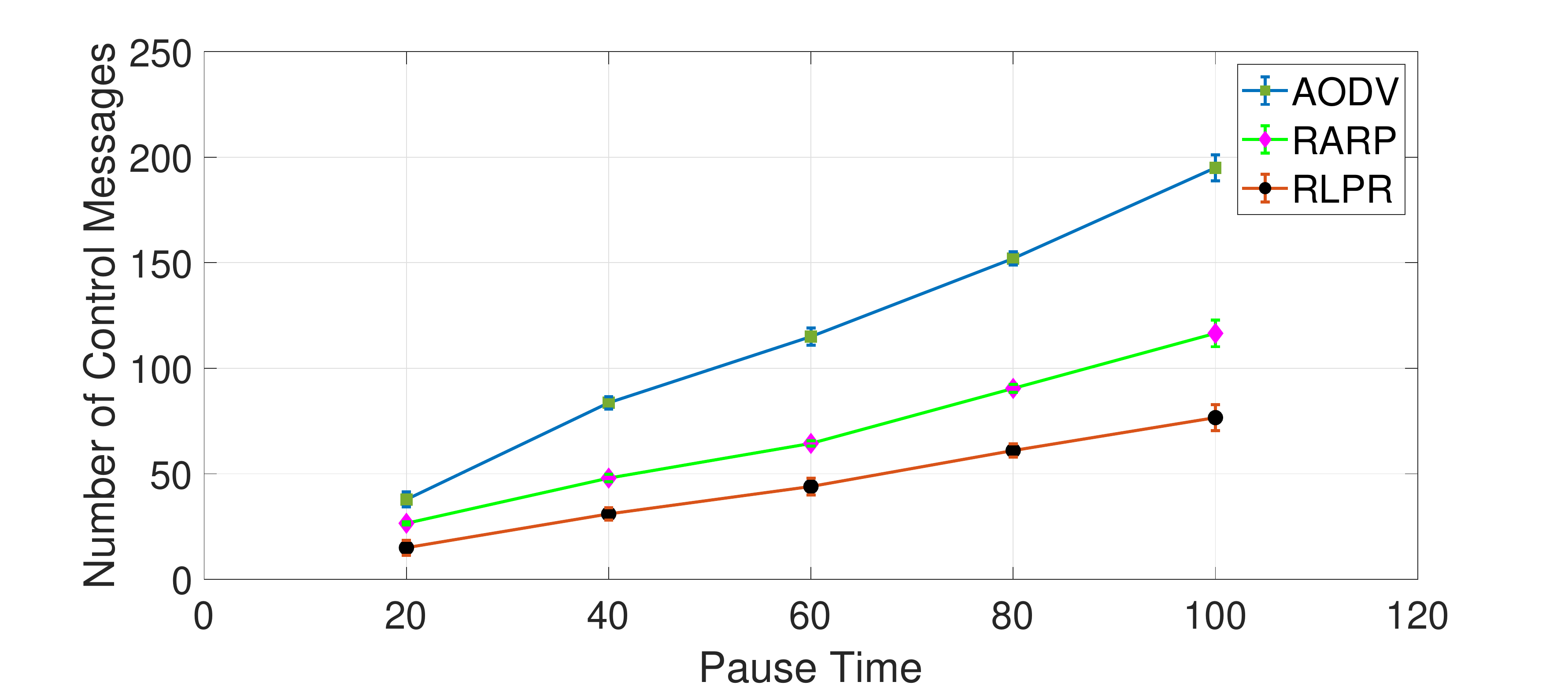}
	\caption{Impact of Pause Time on Control Messages}
	\label{ptcm}
\end{figure}

\begin{figure}[pos=t]
	\centering
	\includegraphics[scale=0.25]{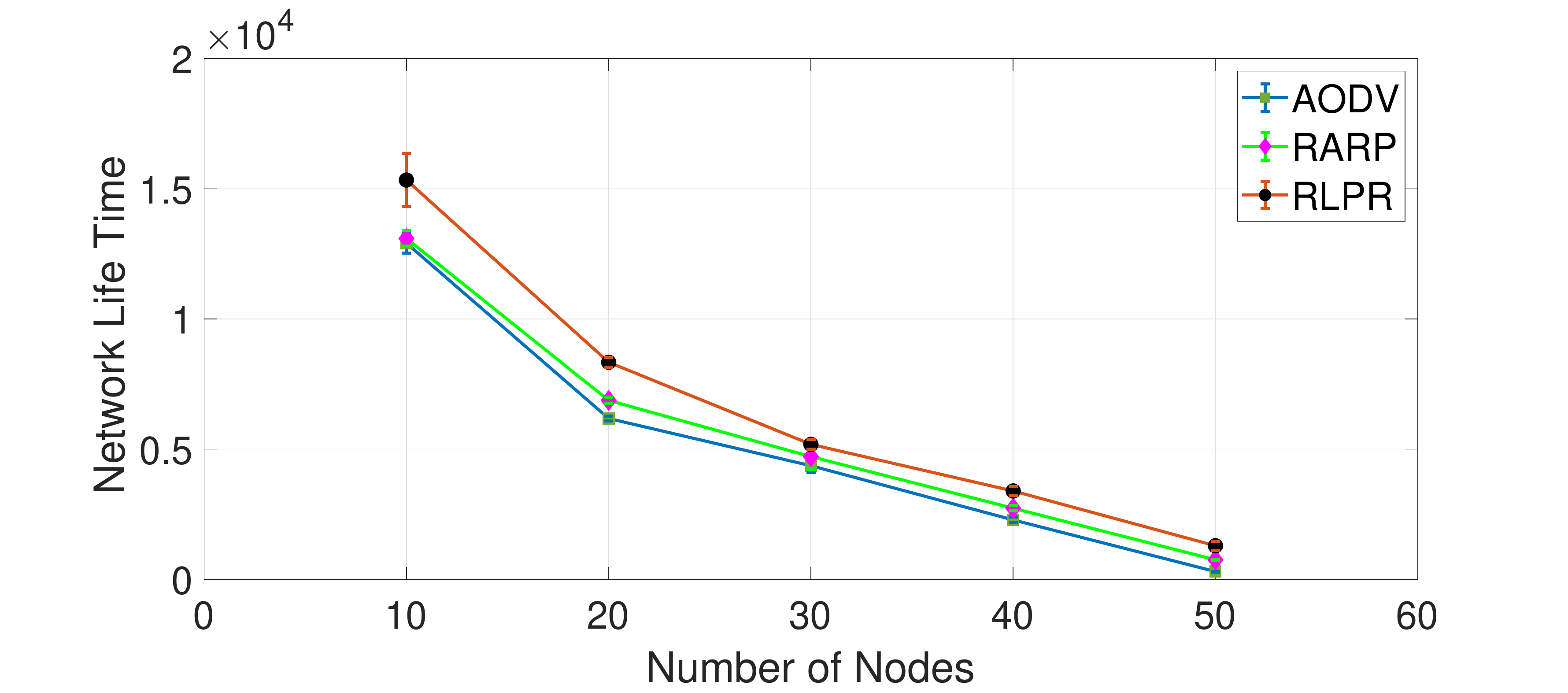}
	\caption{Impact of Number of Nodes on Network Lifetime}
	\label{Nofnodenlt}
\end{figure}

Figure~\ref{ptcm} shows the control message overhead generated in the network by varying the Pause time in a single source and single destination scenario with 20 numbers of nodes in the network. The horizontal axis represents the pause time and the vertical axis shows the corresponding control messages generated by the nodes in the network. As the control message is periodically broadcasted in the network and when the pause time increases more control messages are broadcasted in each protocol. However, RLPR has shown better results as compared to RARP and AODV in terms of control message overhead by varying the pause time. This is because, in RLPR, the route is selected on the basis of composite metric which considered nodes as well as link characteristics. Therefore, with respect to the simulation time, the route does not become a bottleneck and hence the control messages decrease. In contrary to this, RARP and AODV allow all the nodes in the network to re-broadcast the control messages. It is quite possible that multiple routes may contain some common nodes. This may introduce the congestion, which ultimately increases the control messages in the network due to the re-discovery of the path. However, RARP shows better results as compared to AODV, this is because, in RARP, the destination node receives multiple route request packets and among several requests, it selects the best among all the routes. This increases the reliability to some extent which results in a decrease in control messages as compared to AODV. On the other hand, AODV tries to select those nodes in the network which has the path to the destination, without considering that it might be congested. This increases the frequency of failure and increases the control messages overhead.

\begin{figure}[pos=b]
	\centering
	\includegraphics[scale=0.258]{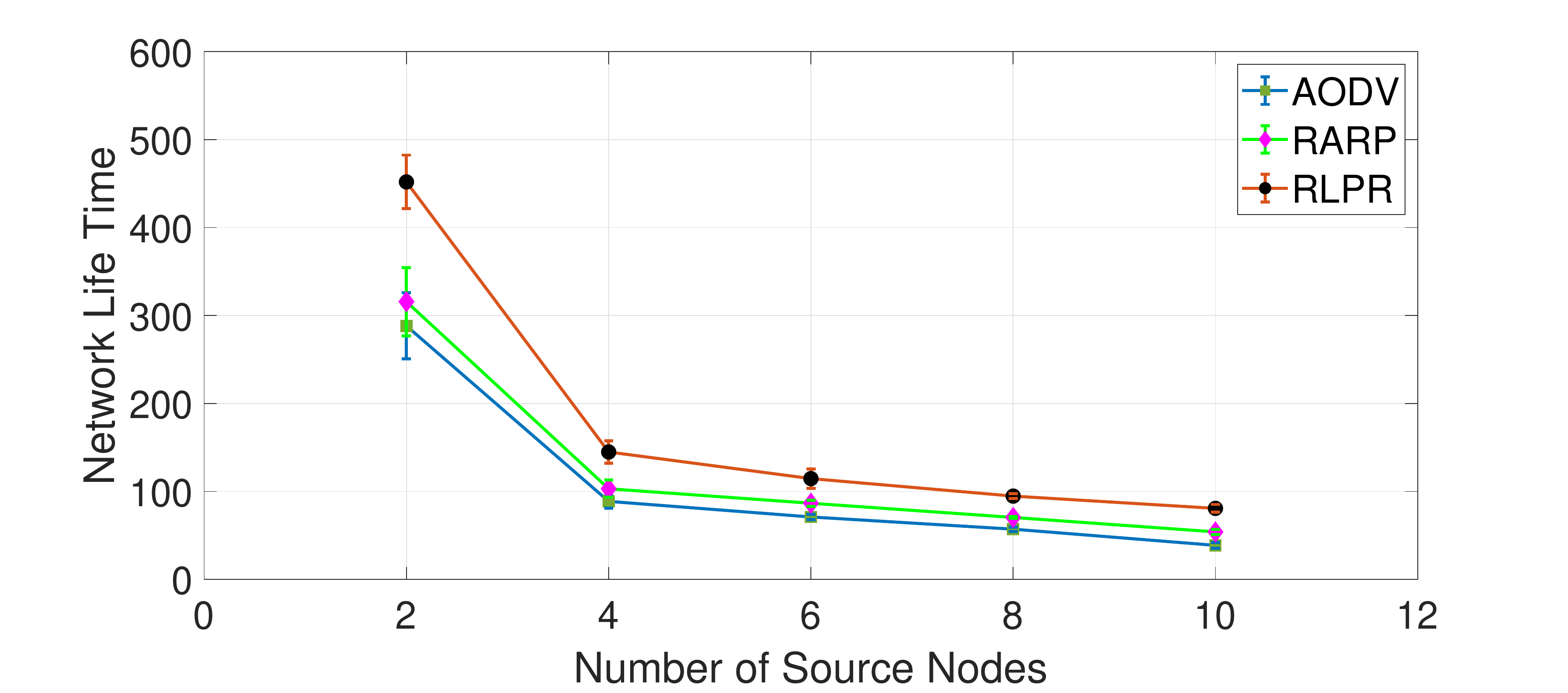}
	\caption{Impact of Number of Sources on Network Lifetime}
	\label{noofsrsnlt}
\end{figure}

Network lifetime is also evaluated and analyzed by varying the number of nodes and by varying the number of sources. Figure ~\ref{Nofnodenlt} shows the network lifetime by varying the number of nodes in a single source and single destination scenario. The horizontal axis represents the numbers of nodes and the vertical axis shows the corresponding network lifetime of nodes in the network. Different energy levels are assigned to all nodes that range from 10 to 100 Joules. RLPR has shown better network lifetime as compared to AODV and RARP. This is because, in RLPR, only those nodes are allowed to rebroadcast the route request whose energy is greater than the threshold value. If the energy of the node is less than 10Jule then that node shall not participate in the communication. Additionally, the nodes along the discovered route shall remain active for a longer period of time because of the front relative selection in the forwarding zone. This selection is based on another criterion that is the composite metric. Based on the composite metric, only that node is allowed to rebroadcast the request whose timer expires first. All other nodes that overhear the request discard the message. In contrary to this, in RARP and in AODV, all the nodes in the network re-broadcast the request, so this decreases the network lifetime, as the energy of the nodes consumes at a high rate because of the shared intermediate node due to the routing criterion. However, RARP shows better results as compared to AODV. This is because, in RARP, nodes are selected based on the better energy level that ultimately enhances the network lifetime as compared to AODV.

\begin{figure}[pos=b]
	\centering
	\includegraphics[scale=0.258]{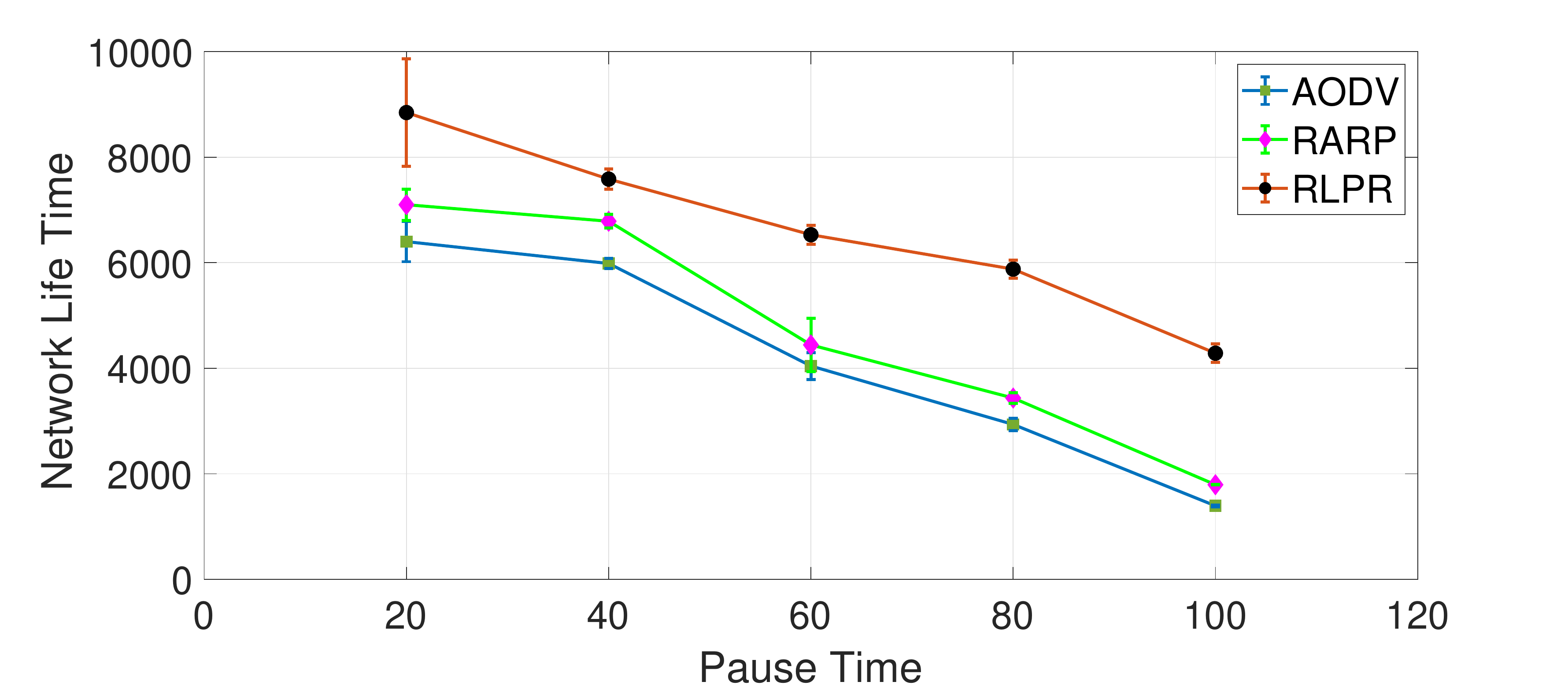}
	\caption{Impact of Pause Time on Network Lifetime}
	\label{ptlt}
\end{figure}

Figure~\ref{noofsrsnlt} shows the network lifetime by varying the number of sources by considering a single destination with 30 numbers of nodes in the network. The horizontal axis represents the numbers of sources and the vertical axis shows the corresponding network lifetime. When the number of sources in the network increases the contention may occur. This is due to the fact that multiple sources in a network try to access the medium or channel, simultaneously, and therefore, the chances of collision increases in the network. Due to collision, nodes consume more energy that significantly decreases the network lifetime. RLPR has shown better network lifetime as compared to AODV and RARP. This is because in RLPR, all source node have their own front relative nodes and there is less chance of collision because in forwarding zone front relative nodes that have high connectivity level and better energy level is allowed to rebroadcast the message in the network. On the other hand, in RARP and in AODV, chances of collision increase, as both of these protocols allow all the nodes to rebroadcast the message in the network, and as a result, nodes consume more energy that significantly decreases the network lifetime.

\begin{figure}[pos=b]
	\centering
	\includegraphics[scale=0.26]{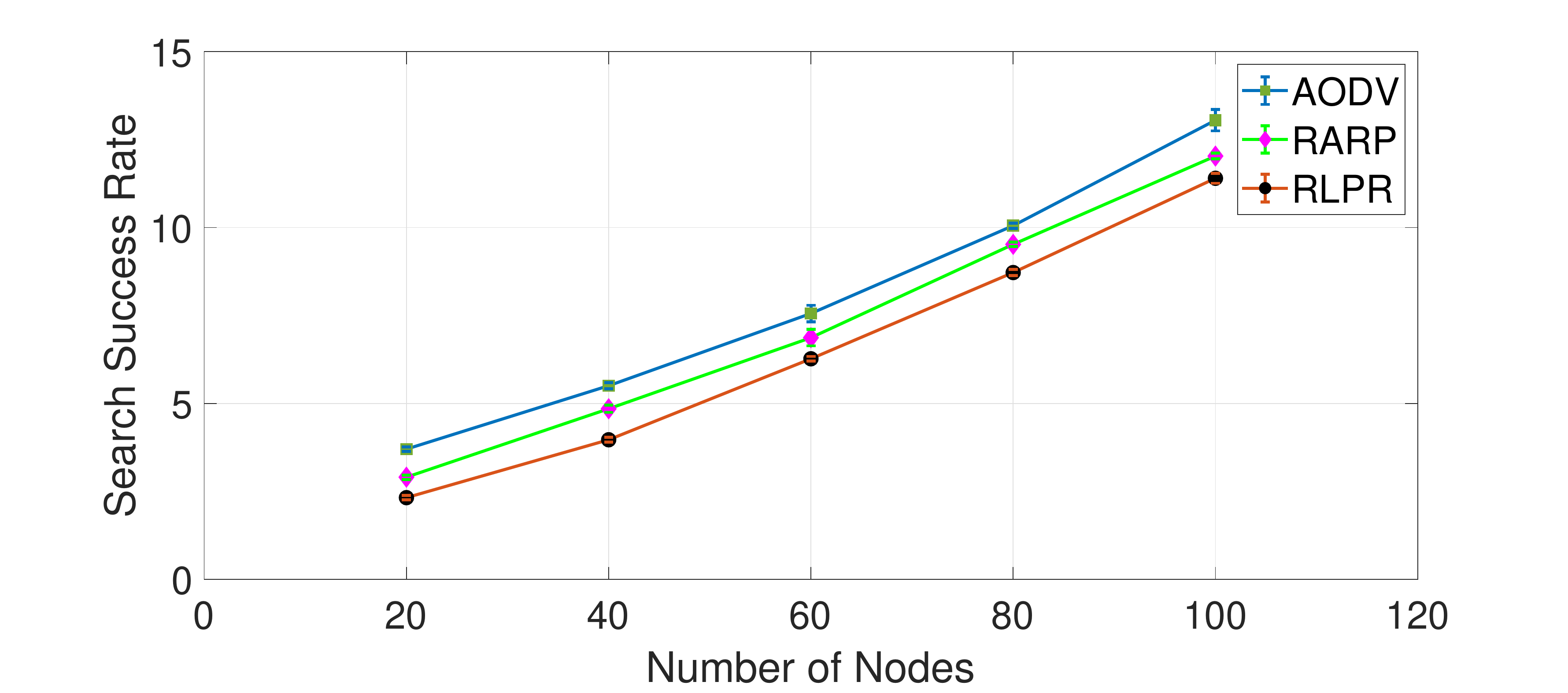}
	\caption{Impact of Number of Nodes on Search Success Rate}
	\label{srt}
\end{figure}

Figure~\ref{ptlt} shows the Network lifetime by varying the Pause time in a single source and single destination scenario with 20 numbers of nodes in the network. The horizontal axis represents the pause time and the vertical axis shows the Network lifetime of the nodes. As control messages are periodically broadcasted in the network, and when the pause time increases, more traffic is generated in the network. Due to this, the contention increases that may increase the packet loss and as a result, the intermediate nodes may become bottleneck nodes that significantly decrease the lifetime. RLPR has shown better network lifetime as compared to AODV and RARP. This is because, in RLPR, the route is discovered by considering the load and energy level of the nodes in the network. Therefore, with the increase in the pause time does not really affect the connectivity level among the nodes. However, in the case of RARP and AODV, the connectivity level decreases that results in an increase in the packet loss and thus the source nodes ultimately rebroadcast the message in the network. This decreases the lifetime of the network. 

Figure \ref{srt} shows the search success rate by varying the number of nodes with a single source and a single destination. Here, the horizontal axis represents the numbers of nodes and the vertical axis shows the search success rate in a network. RLPR has shown better results as compared to RARP and AODV in terms of search success rate. This is because, in RLPR, those nodes are selected which are closer to the destination. With such a criterion, the discovered route contains a smaller number of nodes. Due to the smaller number of nodes, the search success rate decreases. In contrary to this, in RARP and in AODV all the nodes in the network re-broadcast the request to discover the route from source to destination. However, because of the increase in the contention by all the nodes, there is a chance that the request reaches the destination, a bit late. Due to this reason, the search success rate in RARP and in AODV increases. However, RARP shows better results as compared to AODV. This is because, RARP uses the Omni-directional antenna while broadcasting the request and directional transmission towards the predicted location using a directional antenna that enables a longer transmission range, and therefore, the search success rate decreases. On the other hand, AODV shows a higher search success rate as compared to RARP and RLPR. This is due to the fact that AODV just uses the hop count criterion to discover the route from source to destination.

\section{\textbf{Conclusion and future work}}
\label{conclusion}

In the event of major search and rescue operations demand of FANET arises due to unavailable or limited wireless service at critically affected area. However, design of appropriate routing protocol is necessary for successful data communication in FANET. In this research work, we proposed a reliable link adaptive position base routing protocol. Several new features of the RLPR were explained in detailed. Concept of forwarding zone was introduced to select that front relative node which was moving towards destination. High reliability in network was achieved by selecting next relay node with better energy level. Routing path was selected on the basis of composite metric. Performance evaluation of RLPR was carried out by comparing with RARP and conventional AODV routing protocol and simulation was run in different scenarios. The results showed that RLPR increased the network life time and minimized the number of control message that was used for path discovery between source and destination. With different number of nodes in the network the control messages were reduced up to 33\% by RLPR as compared to RARP and 45\% when compared with AODV and life time of the network was enhanced by 55\% when compared with RARP and 65\% when compared with AODV. Also, when the number of sources in the network were increased, RLPR had also shown the reduction in control messages, 49\% as compared to RARP and 57\% when compared with AODV and life time of the network was enhanced by 21\% when compared with RARP and enhanced by 35\% when compared with AODV. Moreover, the search success rate had been determined in RLPR, RARP and in AODV and the results showed that RLPR performed 16\% better as compared to RARP and 28\% as compared to AODV.

The proposed RLPR considered a single destination in the network. In future, the extension can be made by incorporating multiple destination nodes in the network. Additionally, an adaptive forwarding zone using the forwarding angle can be carried out for the selection of a suitable next hop as per the routing criterion. Due to the high mobility, if a node that has a better cost using a composite metric, moves outside the forwarding zone then a source node may increase its forwarding angle and as a result there might be a chance to save more energy. Furthermore, it may also bring more reliability and may able to minimize unwanted and unreliable paths between source and destination.

\bibliographystyle{ieeetr}

\bibliography{mybibfile}

\begin{thebibliography}{10}

\bibitem{bekmezci2013flying}
I.~Bekmezci, O.~K. Sahingoz, and {\c{S}}.~Temel, ``Flying ad-hoc networks
  (fanets): A survey,'' {\em Ad Hoc Networks}, vol.~11, no.~3, pp.~1254--1270,
  2013.

\bibitem{shakoor2019role}
S.~Shakoor, Z.~Kaleem, M.~I. Baig, O.~Chughtai, T.~Q. Duong, and L.~D. Nguyen,
  ``Role of uavs in public safety communications: Energy efficiency
  perspective,'' {\em IEEE Access}, vol.~7, pp.~140665--140679, 2019.

\bibitem{1stPaper}
Z.~{Sheng}, H.~D. {Tuan}, A.~A. {Nasir}, T.~Q. {Duong}, and H.~V. {Poor},
  ``Secure uav-enabled communication using han–kobayashi signaling,'' {\em
  IEEE Transactions on Wireless Communications}, vol.~19, no.~5,
  pp.~2905--2919, 2020.

\bibitem{2ndPaper}
B.~{Wang}, Y.~{Sun}, D.~{Liu}, H.~M. {Nguyen}, and T.~Q. {Duong},
  ``Social-aware uav-assisted mobile crowd sensing in stochastic and dynamic
  environments for disaster relief networks,'' {\em IEEE Transactions on
  Vehicular Technology}, vol.~69, no.~1, pp.~1070--1074, 2020.

\bibitem{3rdPaper}
Z.~{Kaleem}, M.~{Yousaf}, A.~{Qamar}, A.~{Ahmad}, T.~Q. {Duong}, W.~{Choi}, and
  A.~{Jamalipour}, ``Uav-empowered disaster-resilient edge architecture for
  delay-sensitive communication,'' {\em IEEE Network}, vol.~33, no.~6,
  pp.~124--132, 2019.

\bibitem{bekmezci2015flying}
I.~Bekmezci, I.~Sen, and E.~Erkalkan, ``Flying ad hoc networks (fanet) test bed
  implementation,'' in {\em 2015 7th International Conference on Recent
  Advances in Space Technologies (RAST)}, pp.~665--668, IEEE, 2015.

\bibitem{waharte2010supporting}
S.~Waharte and N.~Trigoni, ``Supporting search and rescue operations with
  uavs,'' in {\em 2010 International Conference on Emerging Security
  Technologies}, pp.~142--147, IEEE, 2010.

\bibitem{arafat2019routing}
M.~Y. Arafat and S.~Moh, ``Routing protocols for unmanned aerial vehicle
  networks: A survey,'' {\em IEEE Access}, vol.~7, pp.~99694--99720, 2019.

\bibitem{alshabtat2010low}
A.~I. Alshabtat, L.~Dong, J.~Li, and F.~Yang, ``Low latency routing algorithm
  for unmanned aerial vehicles ad-hoc networks,'' {\em International Journal of
  Electrical and Computer Engineering}, vol.~6, no.~1, pp.~48--54, 2010.

\bibitem{rosati2013speed}
S.~Rosati, K.~Kru{\.z}elecki, L.~Traynard, and B.~R. Mobile, ``Speed-aware
  routing for uav ad-hoc networks,'' in {\em 2013 IEEE Globecom Workshops (GC
  Wkshps)}, pp.~1367--1373, IEEE, 2013.

\bibitem{ni2008performance}
X.~Ni, K.-c. Lan, and R.~Malaney, ``On the performance of expected transmission
  count (etx) for wireless mesh networks,'' in {\em Proceedings of the 3rd
  International Conference on Performance Evaluation Methodologies and Tools},
  p.~77, ICST (Institute for Computer Sciences, Social-Informatics and~…,
  2008.

\bibitem{pu2018link}
C.~Pu, ``Link-quality and traffic-load aware routing for uav ad hoc networks,''
  in {\em 2018 IEEE 4th International Conference on Collaboration and Internet
  Computing (CIC)}, pp.~71--79, IEEE, 2018.

\bibitem{xie2018enhanced}
P.~Xie, ``An enhanced olsr routing protocol based on node link expiration time
  and residual energy in ocean fanets,'' in {\em 2018 24th Asia-Pacific
  Conference on Communications (APCC)}, pp.~598--603, IEEE, 2018.

\bibitem{6913628}
{Yi Zheng}, {Yuwen Wang}, {Zhenzhen Li}, {Li Dong}, {Yu Jiang}, and {Hong
  Zhang}, ``A mobility and load aware olsr routing protocol for uav mobile
  ad-hoc networks,'' in {\em 2014 International Conference on Information and
  Communications Technologies (ICT 2014)}, pp.~1--7, 2014.

\bibitem{forsmann2007time}
J.~H. Forsmann, R.~E. Hiromoto, and J.~Svoboda, ``A time-slotted on-demand
  routing protocol for mobile ad hoc unmanned vehicle systems,'' in {\em
  Unmanned Systems Technology IX}, vol.~6561, p.~65611P, International Society
  for Optics and Photonics, 2007.

\bibitem{hong2019tarcs}
J.~Hong and D.~Zhang, ``Tarcs: A topology change aware-based routing protocol
  choosing scheme of fanets,'' {\em Electronics}, vol.~8, no.~3, p.~274, 2019.

\bibitem{choi2018geolocation}
S.-C. Choi, H.~R. Hussen, J.-H. Park, and J.~Kim, ``Geolocation-based routing
  protocol for flying ad hoc networks (fanets),'' in {\em 2018 Tenth
  International Conference on Ubiquitous and Future Networks (ICUFN)},
  pp.~50--52, IEEE, 2018.

\bibitem{li2017lepr}
X.~Li and J.~Yan, ``Lepr: Link stability estimation-based preemptive routing
  protocol for flying ad hoc networks,'' in {\em 2017 IEEE Symposium on
  Computers and Communications (ISCC)}, pp.~1079--1084, IEEE, 2017.

\bibitem{liu2008clustering}
K.~Liu, J.~Zhang, and T.~Zhang, ``The clustering algorithm of uav networking in
  near-space,'' in {\em 2008 8th International Symposium on Antennas,
  Propagation and EM Theory}, pp.~1550--1553, IEEE, 2008.

\bibitem{zang2011mobility}
C.~Zang and S.~Zang, ``Mobility prediction clustering algorithm for uav
  networking,'' in {\em 2011 IEEE GLOBECOM Workshops (GC Wkshps)},
  pp.~1158--1161, IEEE, 2011.

\bibitem{aadil2018energy}
F.~Aadil, A.~Raza, M.~Khan, M.~Maqsood, I.~Mehmood, and S.~Rho, ``Energy aware
  cluster-based routing in flying ad-hoc networks,'' {\em Sensors}, vol.~18,
  no.~5, p.~1413, 2018.

\bibitem{kuiper2010geographical}
E.~Kuiper and S.~Nadjm-Tehrani, ``Geographical routing with location service in
  intermittently connected manets,'' {\em IEEE Transactions on Vehicular
  Technology}, vol.~60, no.~2, pp.~592--604, 2010.

\bibitem{hyeon2010new}
S.~Hyeon, K.-I. Kim, and S.~Yang, ``A new geographic routing protocol for
  aircraft ad hoc networks,'' in {\em 29th Digital Avionics Systems
  Conference}, pp.~2--E, IEEE, 2010.

\bibitem{gankhuyag2017robust}
G.~Gankhuyag, A.~P. Shrestha, and S.-J. Yoo, ``Robust and reliable predictive
  routing strategy for flying ad-hoc networks,'' {\em IEEE Access}, vol.~5,
  pp.~643--654, 2017.

\bibitem{lin2012novel}
L.~Lin, Q.~Sun, J.~Li, and F.~Yang, ``A novel geographic position mobility
  oriented routing strategy for uavs,'' {\em Journal of Computational
  Information Systems}, vol.~8, no.~2, pp.~709--716, 2012.

\bibitem{broyles2010design}
D.~Broyles, A.~Jabbar, J.~P. Sterbenz, {\em et~al.}, ``Design and analysis of a
  3--d gauss-markov mobility model for highly-dynamic airborne networks,'' in
  {\em Proceedings of the international telemetering conference (ITC), San
  Diego, CA}, pp.~25--28, 2010.

\bibitem{karp2000gpsr}
B.~Karp and H.-T. Kung, ``Gpsr: Greedy perimeter stateless routing for wireless
  networks,'' in {\em Proceedings of the 6th annual international conference on
  Mobile computing and networking}, pp.~243--254, ACM, 2000.

\bibitem{lin2012geographic}
L.~Lin, Q.~Sun, S.~Wang, and F.~Yang, ``A geographic mobility prediction
  routing protocol for ad hoc uav network,'' in {\em 2012 IEEE Globecom
  Workshops}, pp.~1597--1602, IEEE, 2012.

\bibitem{iordanakis2006ad}
M.~Iordanakis, D.~Yannis, K.~Karras, G.~Bogdos, G.~Dilintas, M.~Amirfeiz,
  G.~Colangelo, and S.~Baiotti, ``Ad-hoc routing protocol for aeronautical
  mobile ad-hoc networks,'' in {\em Fifth International Symposium on
  Communication Systems, Networks and Digital Signal Processing (CSNDSP)},
  pp.~1--5, Citeseer, 2006.

\bibitem{marconato2016ieee}
E.~A. Marconato, J.~A. Maxa, D.~F. Pigatto, A.~S. Pinto, N.~Larrieu, and
  K.~R.~C. Branco, ``Ieee 802.11 n vs. ieee 802.15. 4: a study on communication
  qos to provide safe fanets,'' in {\em 2016 46th Annual IEEE/IFIP
  International Conference on Dependable Systems and Networks Workshop
  (DSN-W)}, pp.~184--191, IEEE, 2016.

\bibitem{fabra2017impact}
F.~Fabra, C.~T. Calafate, J.-C. Cano, and P.~Manzoni, ``On the impact of
  inter-uav communications interference in the 2.4 ghz band,'' in {\em 2017
  13th International Wireless Communications and Mobile Computing Conference
  (IWCMC)}, pp.~945--950, IEEE, 2017.

\bibitem{li2015self}
C.~Li, Y.~Chen, X.~Han, and L.~Zhu, ``A self-adaptive and link-aware beaconless
  forwarding protocol for vanets,'' {\em International Journal of Distributed
  Sensor Networks}, vol.~11, no.~8, p.~757269, 2015.

\bibitem{rezende2012virtus}
C.~Rezende, H.~S. Ramos, R.~W. Pazzi, A.~Boukerche, A.~C. Frery, and A.~A.
  Loureiro, ``Virtus: A resilient location-aware video unicast scheme for
  vehicular networks,'' in {\em 2012 IEEE International Conference on
  Communications (ICC)}, pp.~698--702, IEEE, 2012.

\end{thebibliography}

\end{document}